\documentclass{aa}  

\usepackage{graphicx}
\usepackage{txfonts}
\usepackage{hyperref}
\usepackage[utf8]{inputenc}
\usepackage{subfig}
\usepackage{mathrsfs}  
\usepackage{booktabs}
\usepackage{pdflscape}
\usepackage{amsmath}
\usepackage[makeroom]{cancel}
\usepackage{graphicx}
\usepackage{xcolor}
\usepackage{adjustbox}
\usepackage{threeparttablex}
\usepackage{geometry}

\begin{document}

   \title{Bipolar molecular outflow of the very low-mass star Par-Lup3-4}
   \subtitle{Evidence for scaled-down low-mass star formation}
   \author{A. Santamar\'ia-Miranda 
                \inst{1,2,3}
          \and
          I. de Gregorio-Monsalvo\inst{1}
          \and
          N. Hu\'elamo \inst{4}
          \and
          A. L. Plunkett \inst{5}
          \and
          \'A. Ribas\inst{1}
          \and
          F. Comer\'on\inst{6}
          \and
          M.R. Schreiber\inst{2,3}
          \and
          C. L\'opez\inst{7}
          \and
          K. Mu\v{z}i\'{c} \inst{8} 
          \and
          L. Testi \inst{6,9}
          }

   \institute{European Southern Observatory, 3107, Alonso de C\'ordova, Santiago de Chile\\
              \email{asantama@eso.org}
         \and
             Instituto de F\'isica y Astronom\'ia, Universidad de Valpara\'iso, Av. Gran Breta\~na 1111, 5030 Casilla, Valpara\'iso, Chile
                        \and
                        N\'ucleo Milenio Formaci\'on Planetaria - NPF, Universidad de Valpara\'iso, Av. Gran Breta\~na 1111,  Valpara\'iso, Chile 
                        \and
                        Centro de Astrobiolog\'ia (INTA-CSIC); ESAC campus, Camino bajo del Castillo s/n, Urb. Villafranca del Castillo, 28692 Villanueva de la Ca\~nada, Madrid, Spain 
                        \and
                        National Radio Astronomy Observatory, 520 Edgemont Rd, Charlottesville, VA 22903, U.S.A
                \and
                        European Southern Observatory, Karl-Schwarzschild-Strasse 2, 85748 Garching bei M\"unchen, Germany
                        \and
                        Joint ALMA Observatory, 3107 Alonso de C\'ordova, Vitacura, Casilla 19001, Santiago, Chile
                        \and
                        CENTRA, Faculdade de Ci\^{e}ncias, Universidade de Lisboa, Ed. C8, Campo Grande, 1749-016 Lisboa, Portugal
                        \and
                        INAF/Osservatorio Astrofisico di Arcetri, Largo E. Fermi 5, I-50125 Firenze, Italy
                        }

   \date{Received Apr 9, 2020; accepted May 23, 2020}

 
  \abstract
  {Very low-mass stars are known to have jets and outflows, which is indicative of a scaled-down version of low-mass star formation. However, only very few outflows in very low-mass sources are well characterized.}
  {We characterize the bipolar molecular outflow of the very low-mass star Par-Lup3-4, a 0.12 M$_{\odot}$ object known to power an optical jet.}
  {We observed Par-Lup3-4 with ALMA in Bands 6 and 7, detecting both the continuum and CO molecular gas. In particular, we studied three main emission lines: CO(2-1), CO(3-2), and $^{13}$CO(3-2).}
  {Our observations reveal for the first time the base of a bipolar molecular outflow in a very low-mass star, as well as a stream of material moving perpendicular to the primary outflow of this source. The primary outflow morphology is consistent with the previously determined jet orientation and disk inclination. The outflow mass is 9.5$\times$10$^{-7}$M$_{\odot}$ , with an outflow rate of  4.3$\times$10$^{-9}$M$_{\odot}$/yr. A new fitting to the spectral energy distribution suggests that Par-Lup3-4 may be a binary system.}
  {We have characterized Par-Lup3-4 in detail, and its properties are consistent with those reported in other very low-mass sources.  This source provides further evidence that very low-mass sources form as a scaled-down version of low-mass stars.}
  \maketitle


\section{Introduction}
Outflows and jets are ubiquitous structures that accompany the formation of low-mass stars \citep{Arce-2007, Lee20}, especially during
the Class 0 and I evolutionary phases. As the material is accreted from the envelope or the protoplanetary disk onto the protostar, a fraction of the material is expelled as a result of angular momentum conservation. The gas can be ejected in high-velocity collimated jets or in low-velocity disk winds \citep[and references therein]{Hartmann16-1}. 

In recent years, jet-like structures and molecular outflows have been reported in several young very low-mass (VLM) stars and brown dwarfs (BDs), mostly based on optical and infrared spectral and spectro-astrometric observations. 
Examples of well-studied VLM objects with jets are Par-Lup3-4 \citep{FC2005}, LS-R CrA 1 \citep{Whelan09-1}, 2M1207 \citep{Whelan12-1}, ISO 143 \citep{Joergens12-1}, ISO-217 \citep{Joergens12-2}, and ISO-Oph 200 \citep{Whelan18-1}.  

Jets and outflows in VLM stars and BDs have also been detected at centimeter (cm) and millimeter (mm) wavelengths. In the cm regime,  \citet{Morata15-1} discovered compact free-free emission in four sources. In the (sub)mm regime, several VLM stars and BDs have been reported to host jets or outflows in different evolutionary phases: the Class 0/I proto-BDs L1014-IRS \citep{Bourke2005,Huard06-1}, and L1148-IRS \citep{Kauffmann11-1}; and Class II sources ISO-Oph 102 \citep{Phan-Bao08-1}, the VLM star MHO 5 \citep{Phan-Bao11-1}, and GM Tau \citep{Phan-Bao14-1}. Common and notable characteristics of outflows from these sources include a very small physical size (600-1000 au), and a low outflow velocity  (<5 km/s). Other characteristics, such as the ratio between wind mass-loss and accretion rate \citep{Phan-Bao14-1}, have been studied to search for trends among VLM stars and BDs with evolution. The observational uncertainties are still large, however.

\medskip
VLM stars and BDs are very faint and difficult to detect because of sensitivity limitations, resulting in only few studies of mm-wave molecular outflows. \citet{Phan-Bao14-1} detected only three out of eight VLM outflows surveyed in their (sub)mm study. Moreover, there are no sufficient studies (to our knowledge) in the very low-mass regime  on the details at the base of the outflow near the driving source; nor have the inner cavity walls near the launch region been observed and described in great detail. 

The study of outflows from VLM stars may provide further evidence about their formation mechanism. VLM stars may be formed similarly to low-mass stars \citep{Maclow04}, or their formation may be more similar to that of BDs.  The dominant formation mechanism of BDs is still under debate; for a review, see \citet{Joergens14}. Theories such as photoevaporation \citep{WhitworthandZinnecker04-1}, disk fragmentation \citep{Stamatellosandwhitworth09-1}, dynamical ejection \citep{Reipurthandclarke01-1}, or gravoturbulent fragmentation \citep{Padoan2002, Hennebelle2008} can explain the formation of VLM stars and BDs.

In this work, we present new ALMA observations of Par-Lup3-4, a VLM star located in the Lupus 3 molecular cloud. We detect the source in Band 6 and 7 continuum and in three gas emission lines: CO(2-1), CO(3-2), and $^{13}$CO(3-2). These observations reveal that an outflow structure surrounds  Par-Lup3-4. This outflow is consistent with being a scaled-down version of outflows detected in more massive stars. 

\section{Previous observations of Par-Lup3-4}
\label{info}
Par-Lup3-4, located in the Lupus 3 molecular cloud, is a VLM star with a mass of 0.12 M$_{\odot}$ and spectral type M5 \citep{Comeron03-1}. The name of the object is taken from \citealt{Comeron03-1}. The source appears underluminous when compared to similar sources in the same star-forming region, and this is likely due to its edge-on disk orientation. Stellar parameters of Par-Lup3-4 were estimated by fitting its visible and near-infrared spectrum, including a temperature of 3197 K and a luminosity of 0.003 L$_{\odot}$ \citep[][assuming a distance of 200 pc]{alcalaetal14-1} as well as a mass accretion rate of $\log \dot{M}_{\mathrm{acc}}$= -9.1 $\pm$ 0.4 M$_{\odot}$/yr.

A jet from this source was first reported by \citet{FC2005}. A bright knot at 1$.\!\!^{\prime\prime}$3 was detected in H$\alpha$ and [S\,II]. \citet{Comeron2011-1} followed the jet up with narrow-band imaging with the FORS2 instrument, and the knot moved to 2.$\!\!^{\prime\prime}$55 in 7.2 years; it was not detected in H$\alpha$ in the second epoch, and it was fainter in [S\,II] by around 30\,\%. The velocity of the jet is 168 $\pm$ 30 km/s in the plane of the sky, giving a jet inclination of 6$\mathrm{^{\circ}}$7\,$\pm$\,1$\mathrm{^{\circ}}$4. The mass-loss rate was estimated as 3.2 $\times 10^{-10}$M$_{\odot}$/yr \citep{Bacciotti2011}. A detailed study of the optical jet can be found in \citet{Whelan14-1}; the jet extends to $\pm$ 3$^{\prime\prime}$, and is in agreement with the kinematics derived in previous studies. The authors also obtained a better estimate of the ratio $\mathrm{\dot{M}_{out}/\dot{M}_{acc}}$ = 0.05$^{+0.10}_{-0.02}$, which supports theoretical predictions of jet launch \citep{Ferreira2013, Frank2014}.

The early classification as Class I was revisited based on high angular-resolution infrared observations, which revealed no thick envelope around Par-Lup3-4. The spectral energy distribution (SED) was modeled using radiative transfer simulations, and several parameters were estimated. One of them is a disk inclination of 81$\mathrm{^{o}}$, which is compatible with the value obtained by \citet{FC2005}. The maximum derived grain size is $>$10\,$\mu$m, which may be indicative of dust processing. Recently, \citet{Ansdell16-1} and \citet{Ansdell18-1} reported the detection of dust continuum emission from Par-Lup3-4 with ALMA at 335.8 GHz and 225.66 GHz, but there is no report about the bipolar molecular outflow cavity.  

Using data from Gaia DR2, we have estimated the distance to the Lupus 3 cloud to be 155 $\pm$ 10 pc (Santamar\'ia-Miranda, in prep), which agrees within uncertainties with the distance derived by \citet{Zucker20}. This distance is closer than the 200 pc that was previously adopted for this region. We also derived a distance of $\sim$155 pc to Par-Lup3-4.

\section{ALMA observations}
\label{observaciones}
We present ALMA Cycle  3 and 5 observations of Par-Lup3-4 in Bands 6 and 7, respectively. The ALMA Band 6 (1.33 mm) observations were part of a continuum survey that studied the formation mechanisms and evolution of BDs. These observations comprised more than 60 substellar object candidates covering different stages of evolution, from the prestellar core phase to Class II objects (Santamar\'ia-Miranda et al. in prep). Par-Lup-3-4 was included in the list of objects to study, and we present here the continuum emission and CO(2-1) gas emission associated with this source. These observations were performed on 31 March 2016 as part of the Cycle 3 ALMA program 2015.1.00512.S. Data were taken in single-field interferometry mode, and the time on source was 3.5 minutes. The number of antennas used was 43, with minimum and maximum baselines of 15 meters and 452 meters, respectively. The angular resolution achieved was $\sim$0.9 arcsec (see Table \ref{tabla_ppal}, and the largest angular scale was $\sim$11 arcsec.  The field of view was 23 arcsec. Observations were taken with a precipitable water-vapor column of $\sim$1.17 mm. QSO J1517-2422 was used as bandpass and flux calibrator, and QSO J1610-3958 as a phase calibrator.

\begin{table*}[t]

\caption{Dust emission properties derived from ALMA observations }             
\label{tabla_ppal}
\resizebox{\textwidth}{!}{
\begin{tabular}{c c c c c c c c c c}        
\hline\hline                 
Wavelength & Robust  & \multicolumn{3}{c}{Beam size} & rms &  Flux & Peak & Gauss flux& Gauss peak   \\
\cmidrule(lr){3-5}
& & Major axis & Minor axis & PA & & density & intensity & density &  intensity   \\
\noalign{\smallskip}
 [mm] &  &  [arcsec] & [arcsec] & [$^{o}$] & [mJy/beam] &[mJy] & [mJy/beam] & [mJy]  & [mJy/beam]  \\
\hline
1.33  & 2 & 0.93 & 0.84 & 84 & 0.051 & 0.31 & 0.31 & (0.41$\pm$ 0.07)  & 0.28   \\
0.89 & 1.5 & 0.41 & 0.36 & 86 & 0.017 & 0.59 & 0.59 & (0.57 $\pm$ 0.02) & 0.60 \\

\end{tabular}
}
\end{table*}

This project was carried out in dual-polarization mode, and it was designed mainly to detect the continuum and serendipitous gas emission from CO(2-1). The correlator setup included four different basebands, three of them in time-division mode centered at 233.5 GHz, 217.0 GHz, and 219.25 GHz, with a total bandwidth of 1.875 GHz and a spectral resolution of 1.94 MHz. These three spectral windows also covered the frequencies of the C$^{18}$O(2-1), SiO(5-4), and DCN(3-2) transitions with a velocity resolution of $\sim$2.5 km/s, although these molecules were not detected. A fourth baseband was split into two spectral windows with a bandwidth of 468.75 MHz and a velocity resolution of $\sim$0.32 km/s each, centered at 231.15 GHz for continuum detection and at the rest frequency of the CO(2-1) line (230.538 GHz). 

ALMA Band 7 (0.89 mm) observations were performed in Cycle 5 on 24 March 2018 as part of the ALMA program 2017.1.01401.S.  Two individual executions were performed to achieve the requested sensitivity. The time on source was $\sim$72 min. Data were taken using 45 antennas with maximum and minimum baselines of 15 and 783 m, respectively, which provided an angular resolution of $\sim$0.25 arcsec and a largest angular scale of 7.29 arcsec. 
The observations were taken with a precipitable water-vapor column of $\sim$0.60 mm. The field of view  was 16 arcsec.
We observed a single field in dual-polarization mode, dedicating one spectral window of 0.469 GHz bandwidth to observing CO(3-2) with a velocity resolution of 0.11 km/s. The other three spectral windows were selected to study continuum emission with a bandwidth of 1.875 GHz. These three spectral windows covered transitions of $^{13}$CO(3-2), CS(7-6), and SO$_{2}$(4(3,1)-3(2,2)) for a serendipitous detection of gas tracers with a velocity resolution of 0.44 km/s, as well as centering at strategical frequencies to obtain the best atmospheric transmission. Bandpass and flux calibrations were made using QSO J1517-2422, while QSO J1610-3958 was used as the phase calibrator. 

Data were processed using the Common Astronomy Software Applications package (CASA, \citealt{casa}). We used the pipeline version 4.5.3 for the Cycle 3 observations, and version 5.3.0 for Cycle 5 observations. The task CLEAN was used to produce continuum and spectral line images. We used Briggs weighting with different robust parameters in order to obtain the best compromise between spatial resolution and signal-to-noise ratio, varying from 0.5 for CO gas emission lines to 2 for the $^{13}$CO (see Table \ref{tabla_gas} for more details). Primary beam correction was applied before inferring physical parameters from any of the images.

\section{Results}
\label{resultados}
\subsection{Continuum emission} \label{resultados:contin}
ALMA continuum emission images in Band 6 and Band 7 were generated considering all spectral channels from all spectral windows, but excluding channels with spectral line emission. The gas emission lines identified were CO(v=0 2-1) in Band 6, and CO(v=0 3-2) and $^{13}$CO(v=0 3-2) in Band 7. Details of the continuum emission properties are included in Table \ref{tabla_ppal}.

\begin{figure*}[ht]
\includegraphics[width=0.5\textwidth]{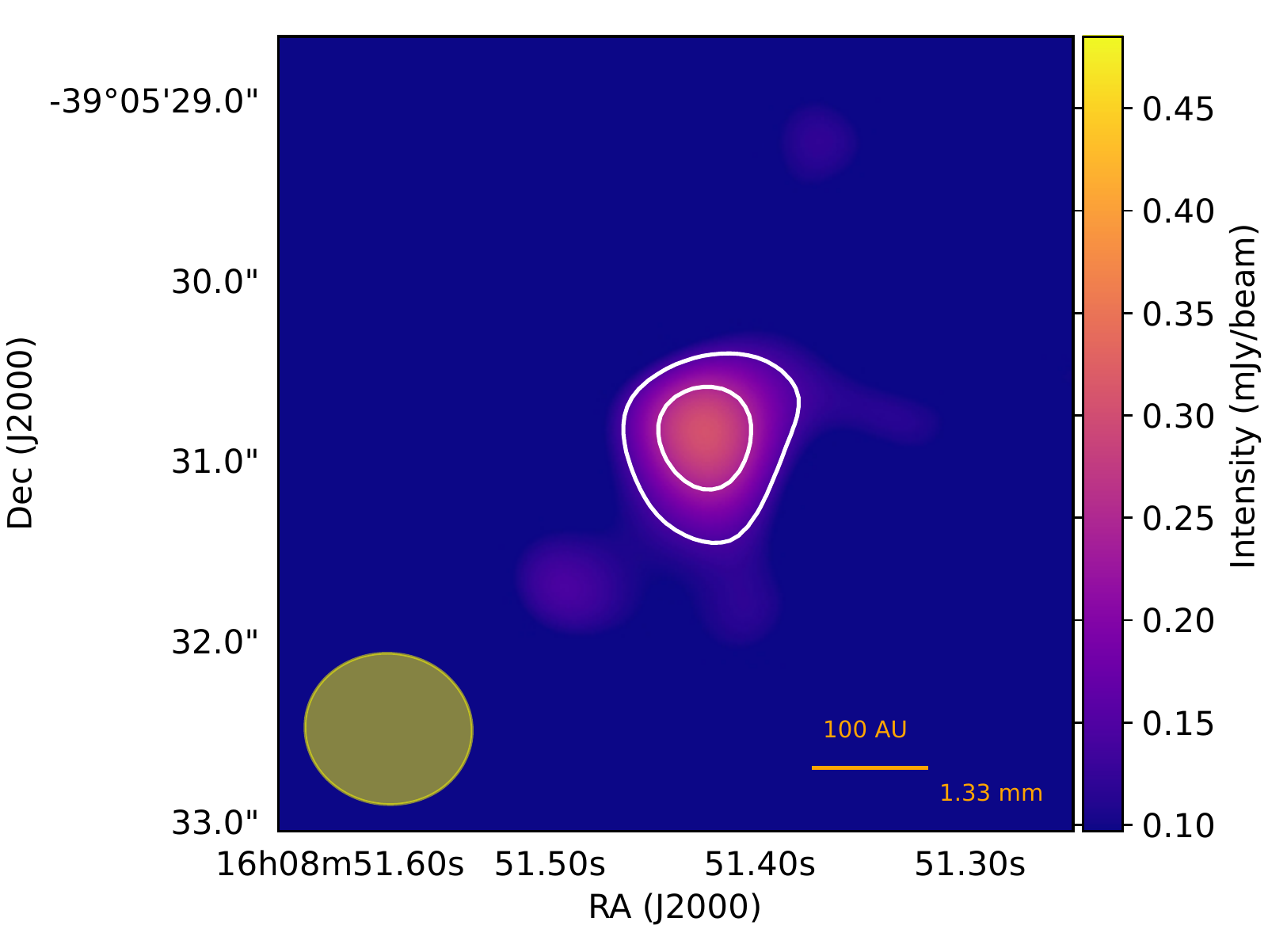}\label{cont_b6}
\includegraphics[width=0.5\textwidth]{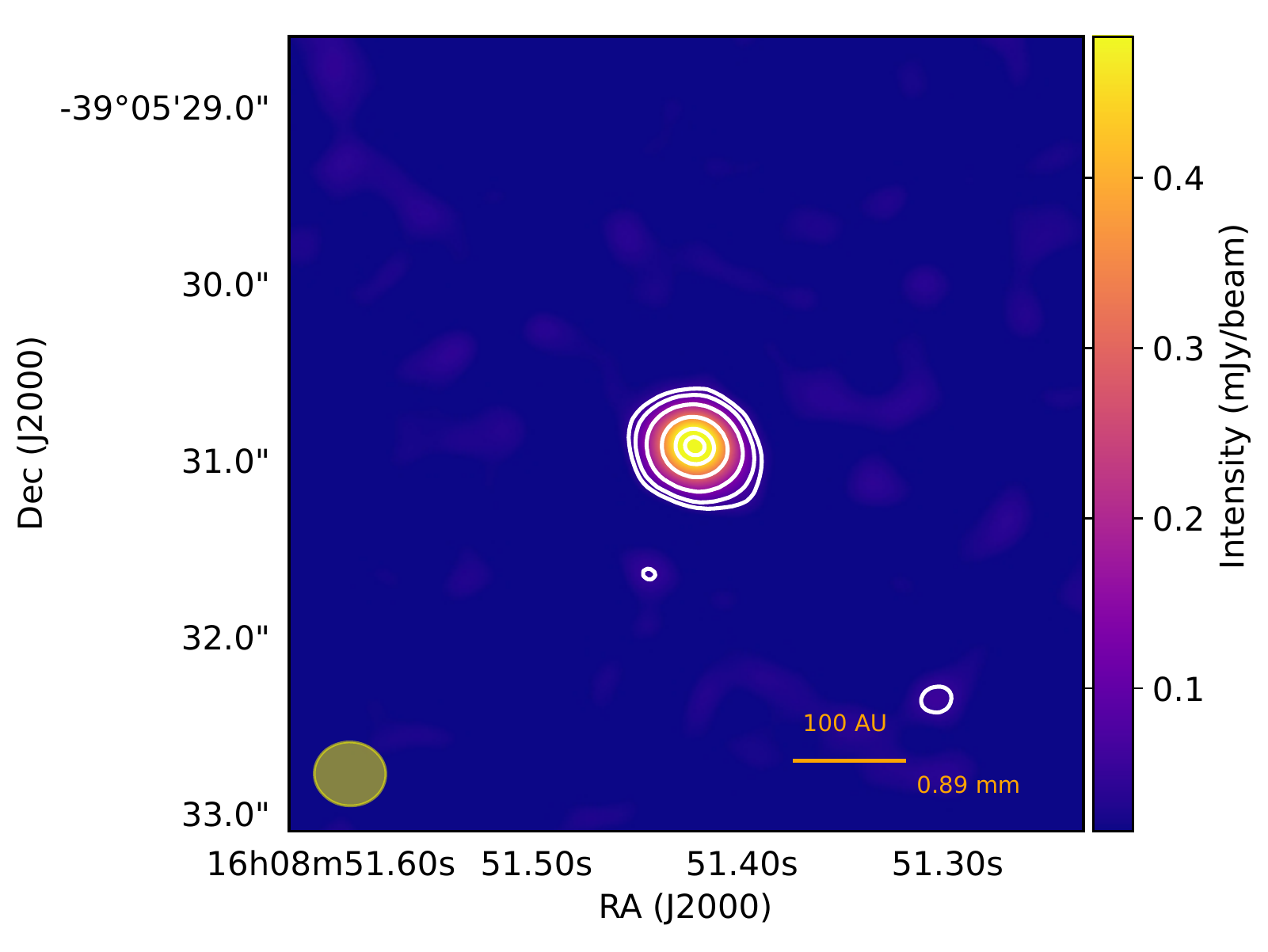}
\caption[ALMA continuum images of Par-Lup3-4 at 1.3 mm]{Left panel: 1.3 mm ALMA continuum image using natural weighting. The white contour scale  is 3 and 5 $\sigma$, where $\sigma$ is the rms noise level of the map. Right panel: 0.89 mm ALMA continuum image using a robust value of 1.5. The white contour scale is 3, 5, 10, 20, 30, and 35 $\sigma$, where $\sigma$ is the rms noise level of the map. The beam size is represented by the yellow ellipse in the bottom left corner in both panels.\label{cont_b7}}
\end{figure*}

We detect dust continuum emission at 1.3 mm (225.27 GHz) at the position R.A.=\,16h08m51.426s, Dec=\,-39$^{\circ}$05'30.82" (see the left panel Fig. \ref{cont_b7}). The ALMA Band 6 continuum image with best sensitivity was obtained using natural weighting (Briggs weighting with robust = 2).  Par-Lup-3-4 was detected at a 6$\sigma$ level with a flux density of 0.31$\pm$0.05 mJy, including a flux calibration error of 10\,\%. The synthesized beam size is $0.93^{\prime\prime}\times0.84^{\prime\prime}$, and the source is spatially unresolved. 
Our results are compatible with previous observations reported by \citet{Ansdell18-1}, who measured a flux of 0.35$\pm$0.11 mJy. 

Dust continuum emission at 0.89 mm (338.15 GHz) is clearly detected at more than 50 $\sigma$ (see right panel Fig.\ref{cont_b7}) at the (J2000) position R.A.=\,16h08m51.424s, Dec=\,-39$^{\circ}$05'30.91". The best-quality image (prioritizing the signal-to-noise ratio while optimizing angular resolution to resolve the spatial structure) was obtained using a robust parameter value of 1.5, yielding an unresolved source with a flux density of 0.59$\pm$0.06 mJy, including a flux calibration error of 10\,\%. \citet{Ansdell16-1} showed the first ALMA continuum image of this source in Band 7, where dust properties were constrained with a flux density of 0.91 $\pm$ 0.26 mJy. Our results are marginally compatible with theirs within 1$\sigma$ uncertainty. In an attempt to improve the spatial resolution, we generated an image using uniform weighting, but still did not resolve the source, we therefore additionally adopted the value of the synthesized beam ($\sim$0.24 arcsec) of the image with uniform weighting as an upper limit of the dust disk size (60 au diameter at 155 pc).

Assuming the ALMA continuum emission comes from thermal dust emission, and considering optically thin emission, we derive the total dust mass from \citet{Hildebrand1983} as
\begin{equation}
      M =  \frac{S_{\lambda} D^{2}}{B_{\lambda}(T_{dust}) \kappa_{\lambda}},
\end{equation}
where $S_{\lambda}$  is the flux density from Table \ref{tabla_ppal}, $D$ is the distance to the source (155\,pc), and $B_{\lambda}(T_{dust})$ is the Planck function at a temperature T$\mathrm{_{dust}}$. The temperature was derived using T$_{\mathrm{dust}}$\,=\,25\,K \,$\times$\,(L$_{*}$/L$_{\sun}$)$^{0.25}$ \citep{Andrewsetal13-1}. We used the stellar radius to obtain the luminosity (L$_{*}$), assuming an effective temperature of 3197 K \citep{alcalaetal14-1}. We did not use the standard L$_{*}$ estimates  (i.e., visual magnitude plus bolometric correction) because the extinction toward the photosphere is highly uncertain because the source is seen edge-on and therefore appears underluminous in the Hertzsprung-Russell diagram. We first used a radius of 1.1 R$_{\sun}$ \citep{Huelamo10-1}, obtaining a temperature of 15 K. Then, we used a radius of 2 R$_{\sun}$, as the result of our SED fitting (see Section \ref{fiteo}), obtaining a temperature of 20 K. $\kappa_{\lambda}$ is the absorption coefficient obtained from \citet{Ose94} for thin ice mantles and a density of $10^{6}$ cm$^{-3}$. We interpolated for the wavelength of 1.3 mm and for 0.89 mm and obtained values of $\kappa$= 0.85  cm${^2}$g${^{-1}}$  and  1.8 cm${^2}$g${^{-1}}$, respectively. The derived  masses for the dust at a temperature of 20 K are 0.28  $\pm$ 0.05 M$\mathrm{_{\oplus}}$ for Band 7 and 0.60 $\pm$ 0.15 M$\mathrm{_{\oplus}}$ for Band 6. Using a temperature of 15 K, we obtained a dust disk mass of 0.43 $\pm$ 0.08 M$\mathrm{_{\oplus}}$ for Band 7 and 0.88 $\pm$ 0.22 M$\mathrm{_{\oplus}}$ for Band 6. When the fluxes from  \citet{Ansdell16-1} and \citet{Ansdell18-1} with a distance of 155 pc instead of their assumed 200 pc, and the opacity law from from \citet{Ose94} are used, our results are compatible with theirs within the errors.

\subsection{Gas emission lines}
We detected three different gas emission lines toward this source for the first time with ALMA. In this section we describe structures with 3 $\sigma$ detection or more. The results for  CO(3-2), CO(2-1), and $^{13}$CO(3-2) are summarized in Table \ref{tabla_gas}, and they are described in detail in the following subsections.

\subsubsection{CO(3-2)}
\label{sec:co_3-2_descripcion}
\begin{figure*}

\includegraphics[width=0.5\textwidth]{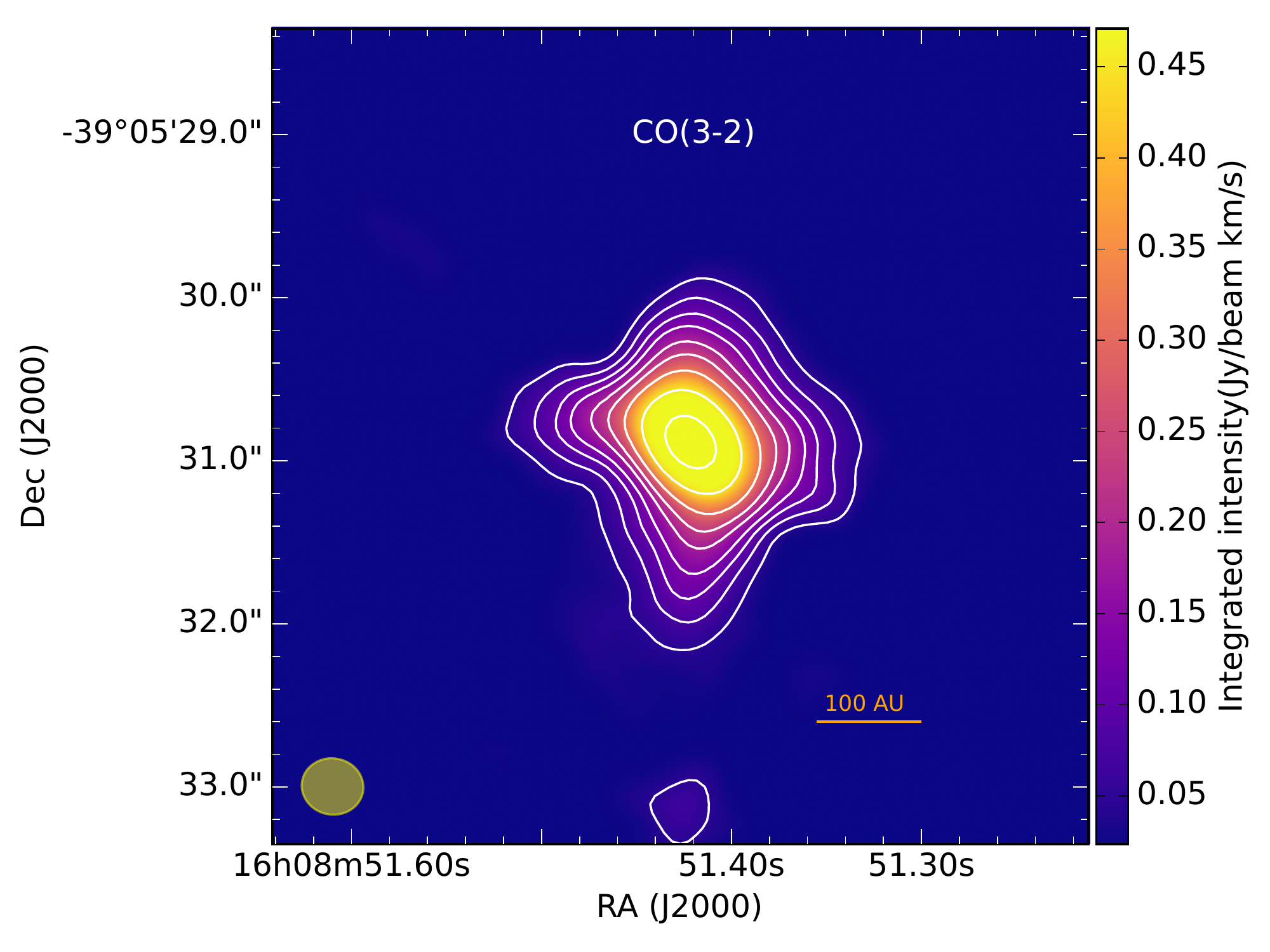}
\includegraphics[width=0.5\textwidth]{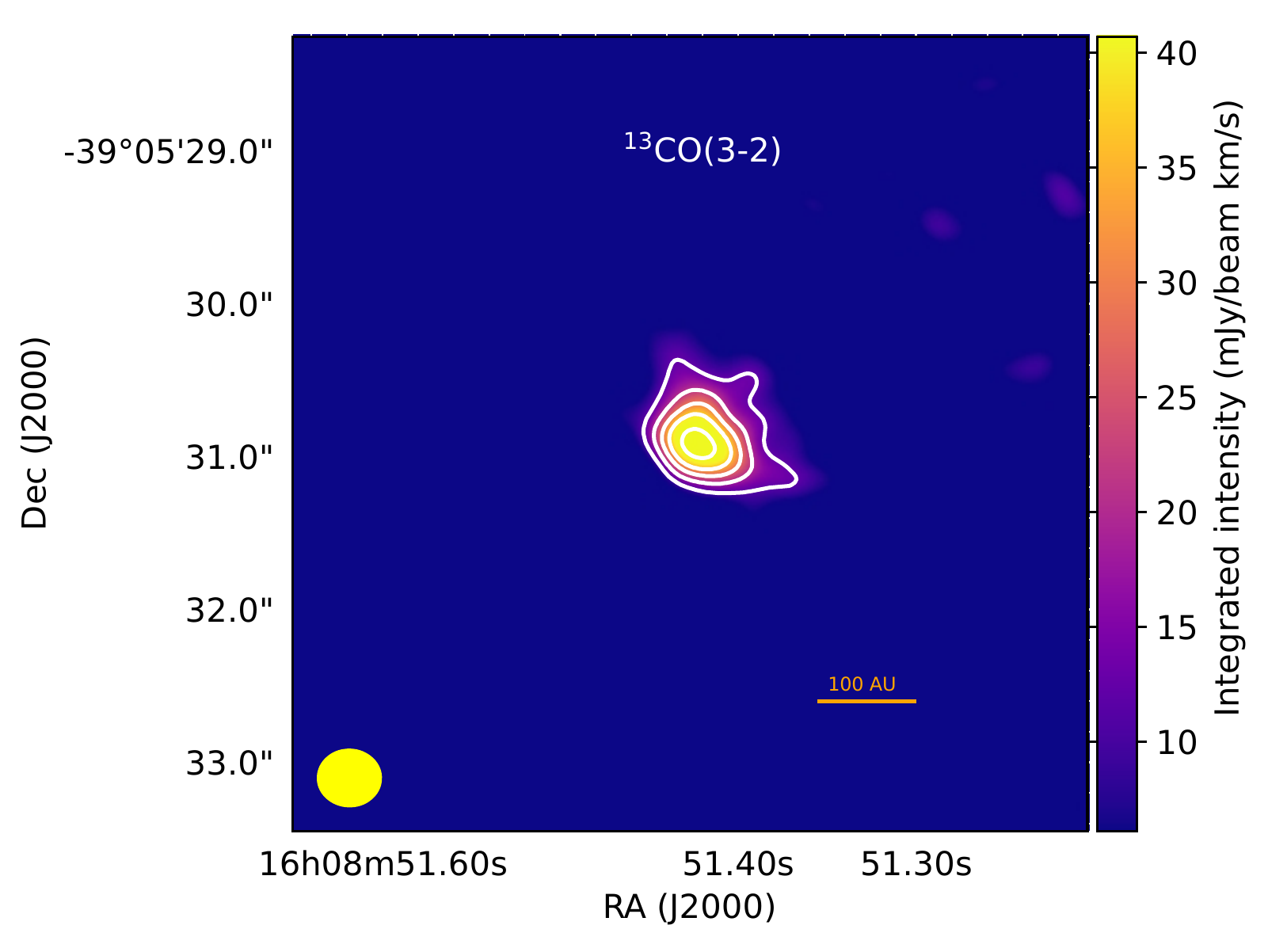}
\begin{center}
\includegraphics[width=0.5\textwidth]{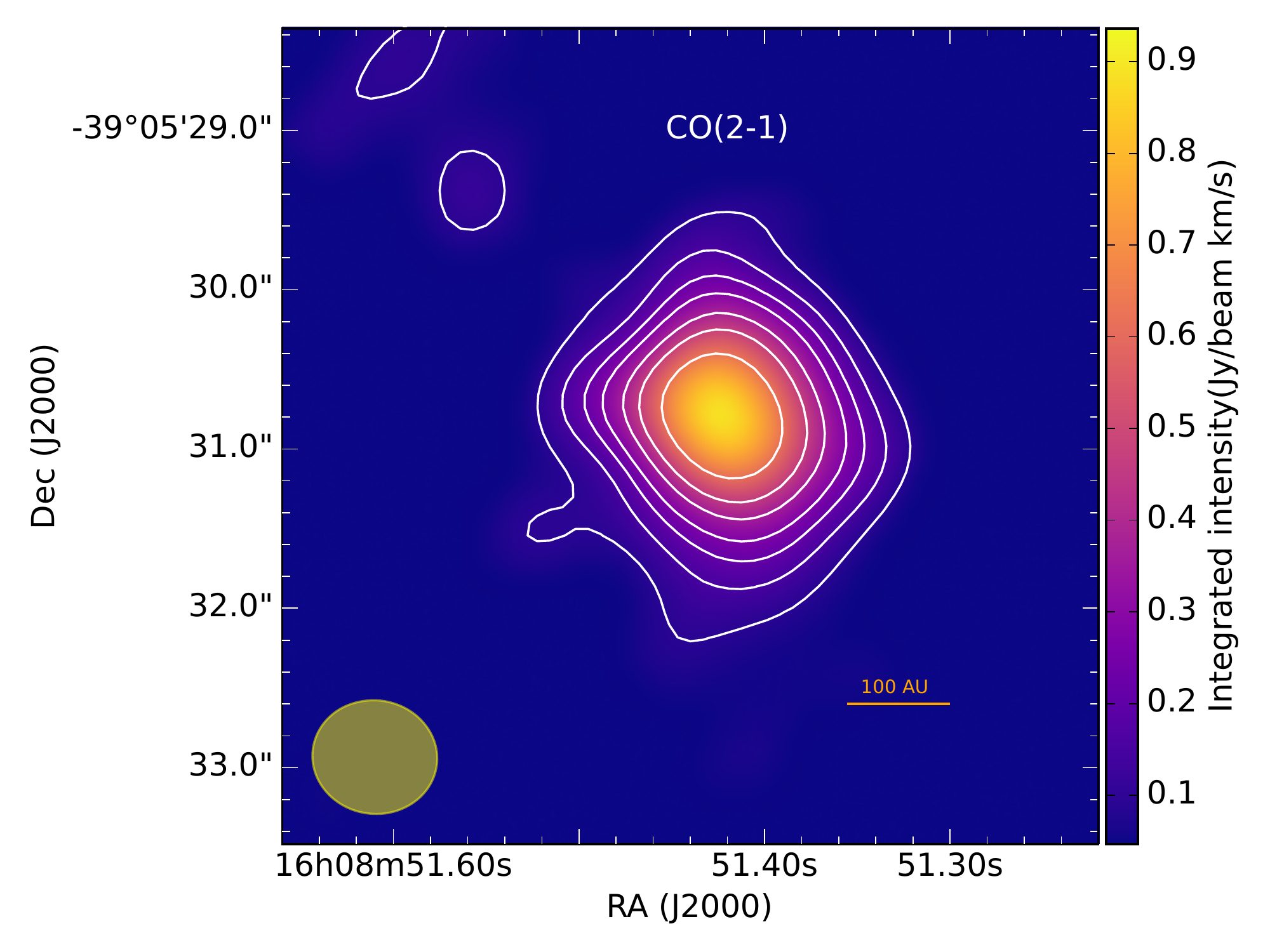}
\end{center}
\caption[ALMA gas images of Par-Lup3-4 at 1.3 mm]{Top left panel: CO(3-2) flux-integrated ALMA map from velocity -3 to 11 km/s with a robust value of 1. The white contours are 3, 5, 7, 9, 12, 15, 20, 30, and 50 times the rms. Top right panel: $^{13}$CO flux-integrated ALMA map from velocity -0.08 to 7.44 km/s. The white contours are 3, 5, 7, 9, and 12  times the rms. Bottom panel: CO(2-1) flux-integrated ALMA map from velocity -2.62 to 10.08 km/s. The white contours are 3, 5, 7, 9, 12, 15, and 20 times the rms. The beam size is represented by a yellow ellipse in the bottom left corner of the three panels. The three panels show a zoom-in of the main central core region. A zoom-out image can be found in Fig. \ref{channel_map_CO2-1}, Fig. \ref{channel_map_CO3-2}, and Fig. \ref{channel_map_13CO}  for CO(2-1),  CO(3-2), and $^{13}$CO,  respectively.} 
\label{figure:zero_images_b}
\end{figure*}

\begin{figure*}
\newgeometry{inner=2.5cm, outer=2.5cm}
\includegraphics[height=1.05\textheight]{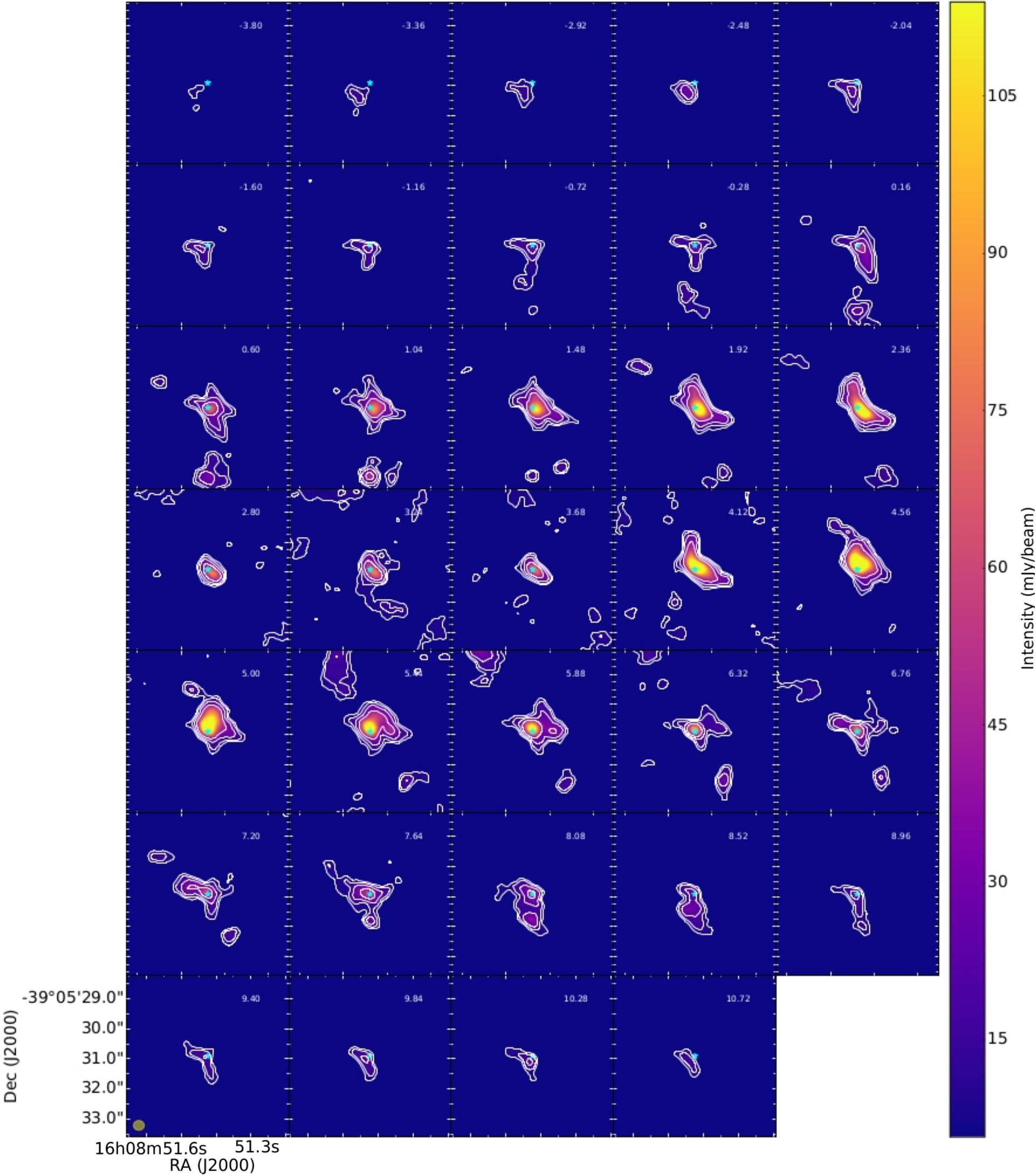}
      \caption[Zoom out ALMA CO(3-2) channel emission map of Par-Lup3-4]{ \label{channel_map_CO3-2} Zoom-out CO(3-2) channel maps toward Par-Lup3-4 using robust value of 1. We used a range >50 k$\lambda$ to eliminate the extended emission. We binned the image to a velocity resolution of 0.44 km/s. The velocity of the channels is shown in the LSR frame in km/s, centered at the frequency of CO(3-2). All maps share the same linear color scale.  White contour levels are 3, 5, 9, and 17 $\sigma$. $\sigma$ is the rms noise level of the map. The cyan star marks the position of the peak intensity in the continuum image. The beam size is represented by a yellow ellipse in the bottom left corner.}
\restoregeometry
\end{figure*}

\begin{figure}
\includegraphics[width=\hsize]{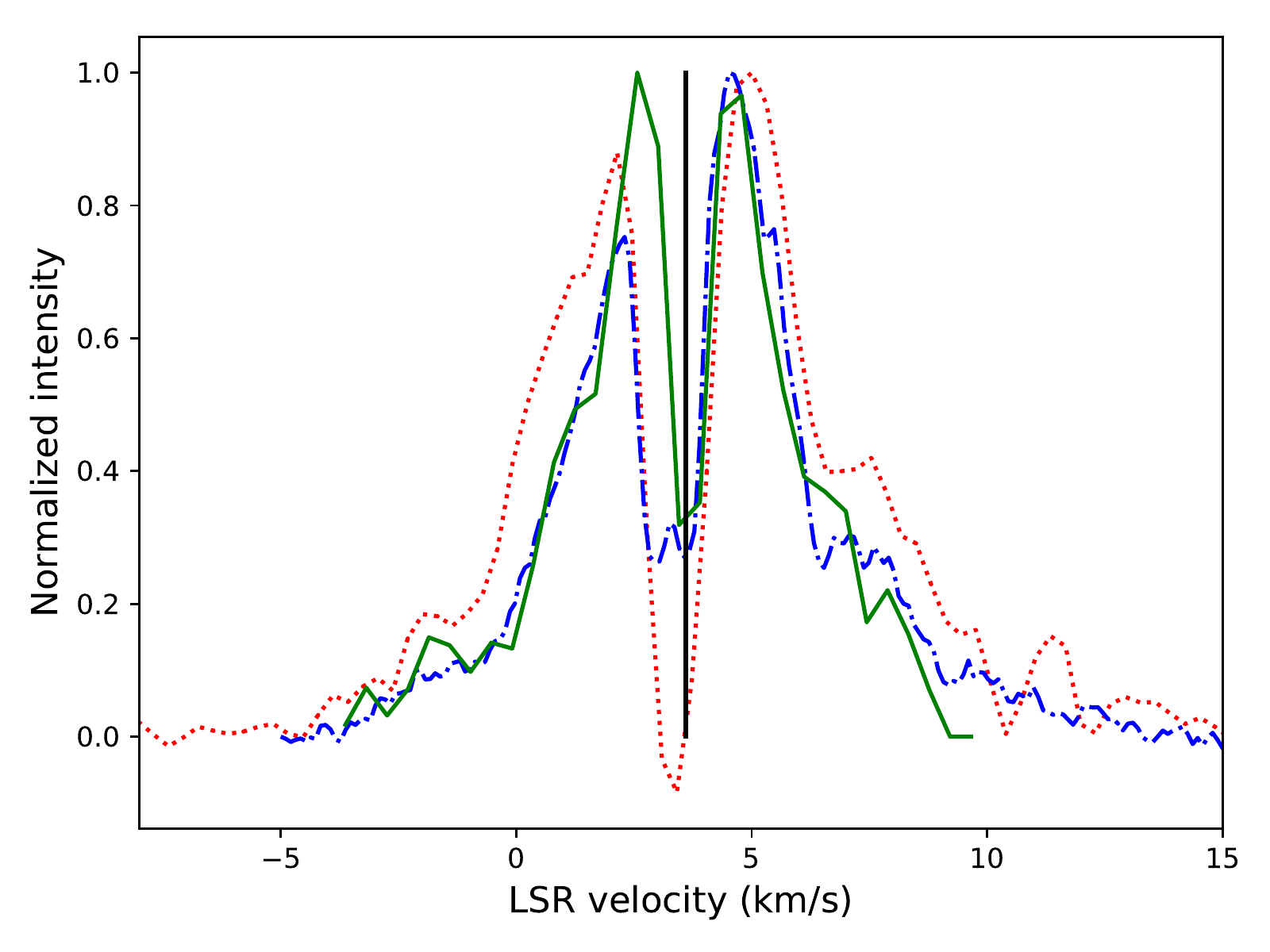}
 \caption[$^{13}$CO, CO(2-1) and CO(3-2) spectra]{$^{13}$CO (solid green), CO(2-1) (dashed red), and CO(3-2) (semidashed blue) spectra averaged over the 5\,$\sigma$ contour level in the integrated flux maps for each line applied to the spectral cube and centered on Par-Lup3-4. The black line represents the system velocity of Par-Lup3-4, i.e., the velocity average of gas, obtained from  $^{13}$CO, $\sim $3.7km/s. Blueshifted material is lower than 3.45 km/s and redshifted material is greater than 3.90 km/s.}

\label{espectro_central}
\end{figure}

CO(3-2) (top left panel in Fig. \ref{figure:zero_images_b}) emits in a velocity interval between -2.92 to 11.60 km/s (see Fig. \ref{channel_map_CO3-2}). The spectrum of Par-Lup3-4 displays a double-peak profile with blueshifted and redshifted wings and a self-absorption feature at $\sim$3.4 km/s (see Fig. \ref{espectro_central}), probably due to the cold foreground parental molecular cloud. The self-absorption feature is close to the source systemic velocity at $\sim$3.7 km/s (see Sect. \ref{13co}).
The CO(3-2) blueshifted emission spans velocities between -2.92 to 2.80 km/s, and the redshifted emission is between 3.70 to 10.72 km. The blueshifted arc-like structures are seen from -2.92 to 2.36 km/s in the southeast and from 0.60 to 2.80 km/s to the northwest. The redshifted emission comes from a northwest arc-like structure that emits between 3.68 to 6.32 km/s and from a southeast structure that extends from 5.00 to 10.72 km/s. Integrated red- and blueshifted CO(2-1) contour maps are provided in Appendix \ref{red_blue_b7}

\begin{table*}[!hbt]

\caption{Gas properties of the ALMA detection}             
\label{tabla_gas}
\begin{tabular}{c c c c c c c c}        
\hline\hline                 
Molecular & Robust & \multicolumn{3}{c}{Beam size} &  rms &  Integrated & Peak   \\
\cmidrule(lr){2-5}
transition & &  Major axis & Minor axis & PA & & intensity\footnote{1} & intensity  \\
\noalign{\smallskip}
&  & [arcsec] & [arcsec] & [$^{o}$]  & [Jy/beam km/s] & [Jy km/s ] & [Jy/beam]  \\
\hline
CO(3-2) & 1 &  0.38 & 0.35 & 80 & 1.57$\times$10$^{-2}$ & 3.79 & 0.99    \\
CO(2-1) & 1 &  0.79 & 0.71 & 82  & 3.12$\times$10$^{-2}$ & 1.97 & 0.89 \\
$^{13}$CO(3-2)& 2 & 0.41 & 0.37 & 87 & 4.07$\times$10$^{-3}$ & 8.16$\times$10$^{-2}$ & 5.50$\times$10$^{-2}$  \\
\end{tabular}
\begin{tablenotes}
\begin{footnotesize}
\item[1] $^{1}$Obtained over a 3$\sigma$ contour that corresponds to an area of $\sim$3.8 arcsec$^{2}$, $\sim$5.0 arcsec$^{2}$ , and $\sim$0.7 arcsec$^{2}$ for CO(3-2), CO(2-1), and $^{13}$CO(3-2), respectively.
\end{footnotesize}
\end{tablenotes}

\end{table*}

\begin{figure}
\centering
\includegraphics[width=\hsize]{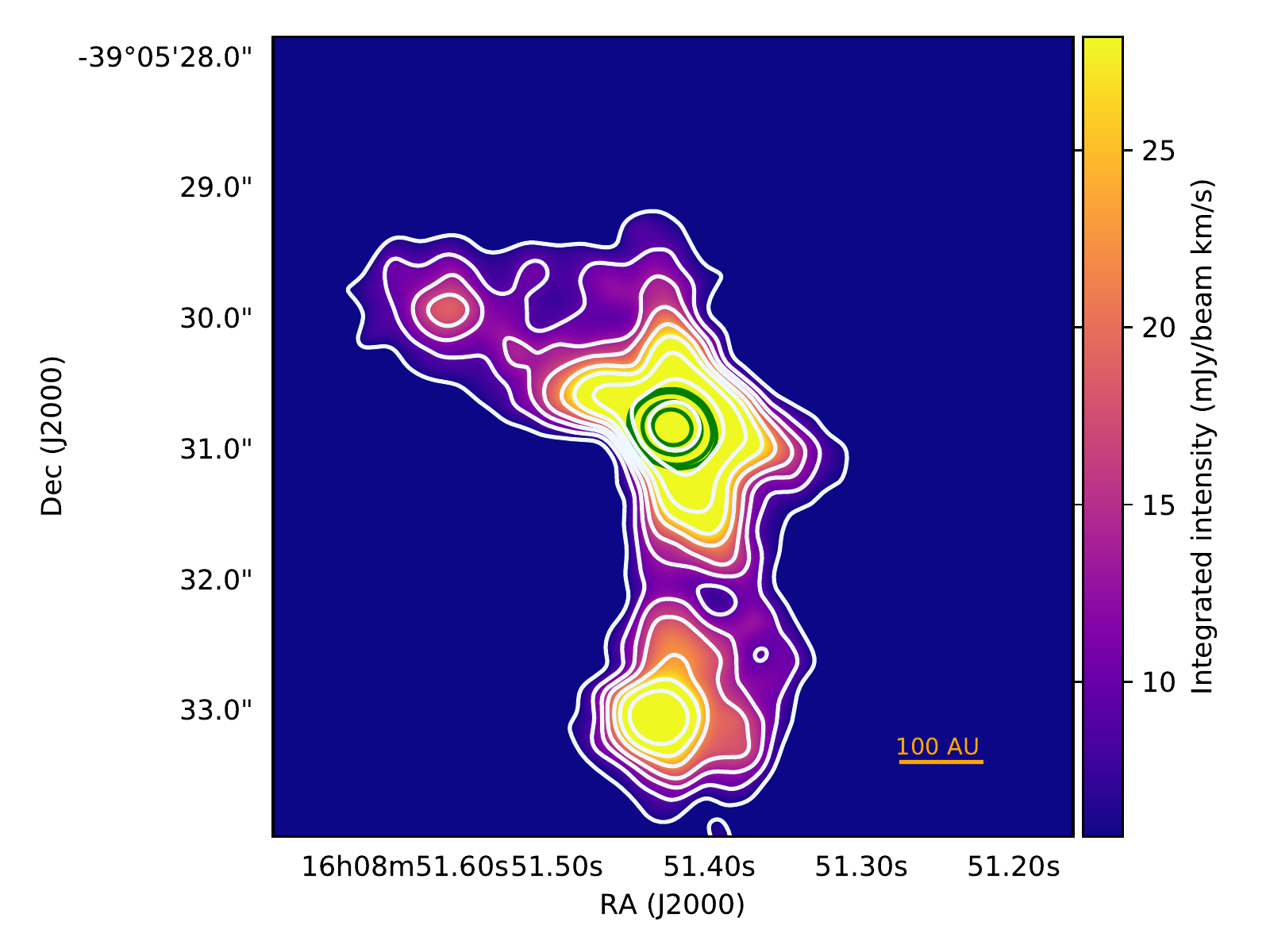}
      \caption[Jet and left bipolar outflow cavity of Par-Lup3-4]{CO(3-2) flux-integrated map in color to show the bipolar molecular outflow cavity. Only channels that show bipolar emission between [-3 km/s, 2.25 km/s] and [6.75 km/s, 10.25km/s]) were included. We excluded channels with possible cloud contamination. White contour levels are 3, 5, 7, 9, 12, 15, 20, and 30 times the rms with a robust value of 2. Green contours are the ALMA continuum image at 0.89 mm using 5, 7, 15, and 25 times the rms value with a robust value of 1.5. $\sigma$ is the rms noise level of the map. Only channels with a bipolar molecular structure were chosen.}
\label{jet}
\end{figure}

CO(3-2) traces low-velocity outflowing material with an inclination near the plane of the sky, as revealed by the different arc-like quasi-symmetric structures with superimposed blue- and redshifted emission that traces the base of a compact bipolar outflow very close to the position of Par-Lup3-4 (Fig. \ref{jet}).  This outflow has the same orientation as the jet and the counterjet detected by \citet{FC2005}. The CO(3-2) lobe structures clearly delineate the southeast and northwest side of the outflow cavities that result from the interaction between the ejected material with the surrounding envelope.

Cloud emission is seen as an inhomogeneous distribution of material that is spread randomly throughout the whole map between 2.51 km/s to 4.41 km/s and between 5.26 km/s  and 5.58 km/s. Beyond 5.90 km/s and below 7.20 km/s, we cannot distinguish clearly between cloud emission and outflow emission.

\begin{figure*}
\includegraphics[width=0.53\textwidth]{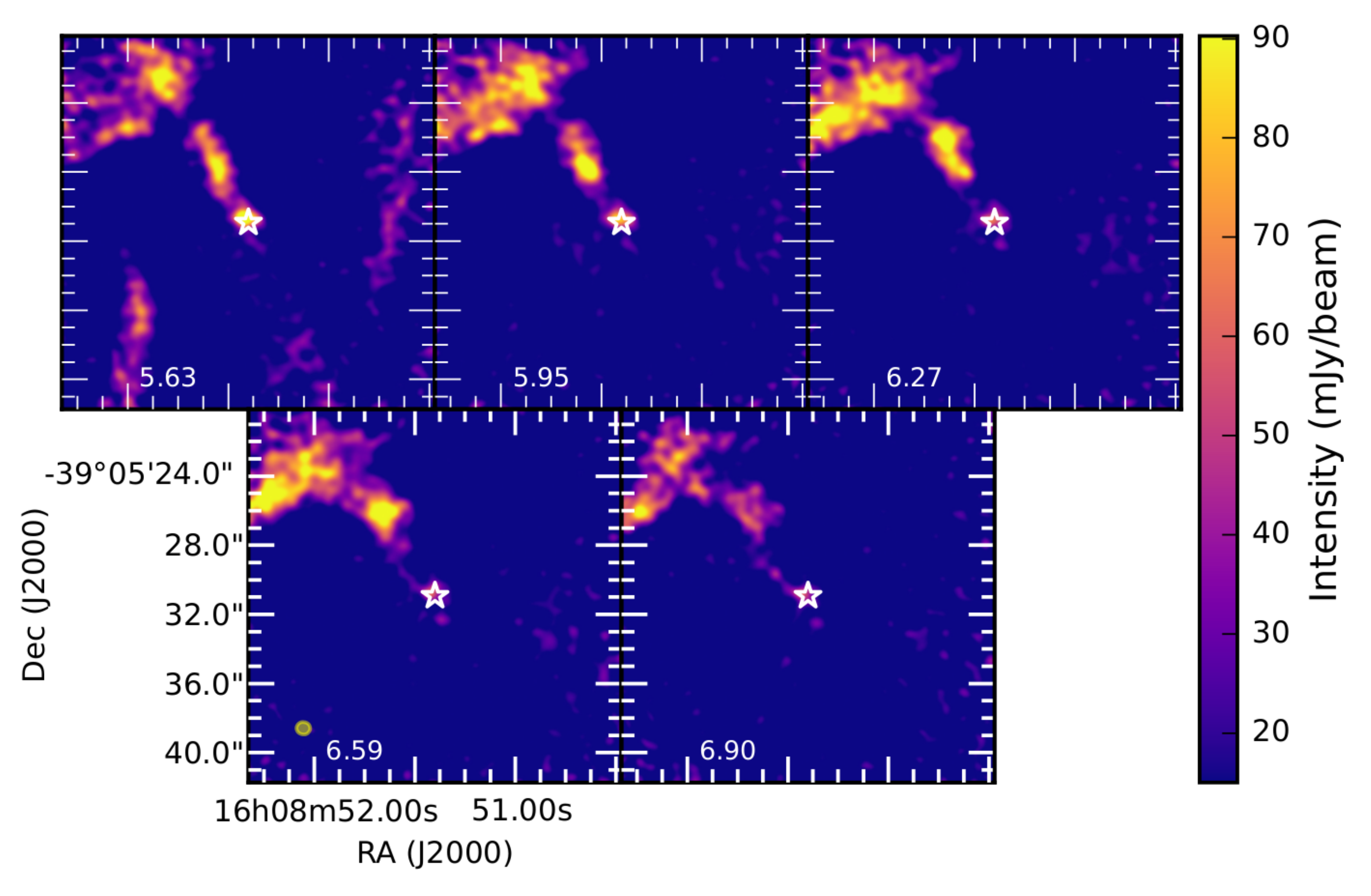}
\includegraphics[width=0.46\textwidth]{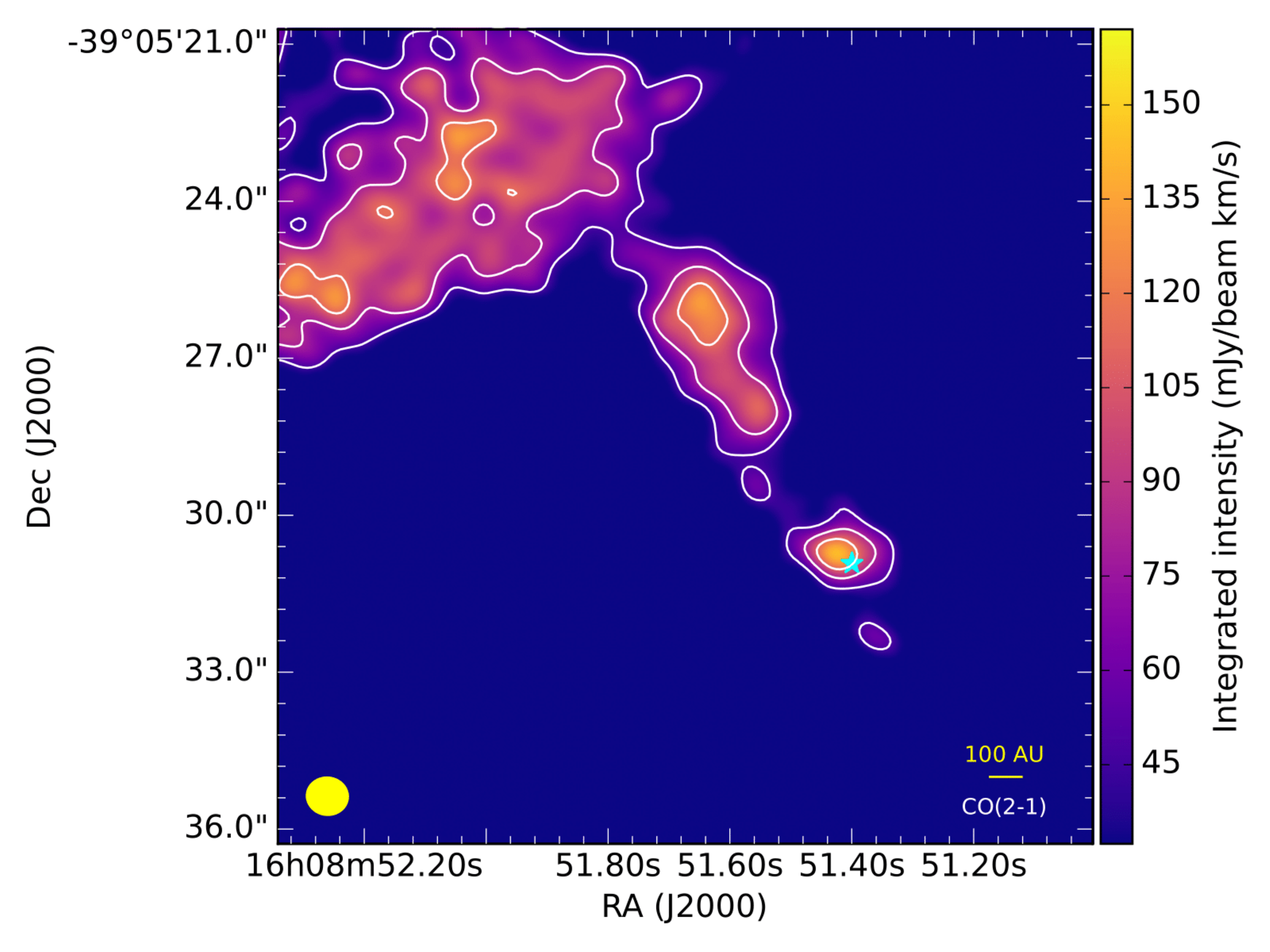}

\caption{Left panel: CO(2-1) ALMA channel maps toward Par-Lup3-4 and the possible secondary outflow following the northeast direction.  All the maps share the same linear color scale with a robust value of 1.  Right panel: ALMA-integrated emission of CO(2-1) from a velocity of 5.63 to 6.90 km/s. Contours show 3, 5, and 7 times the rms (6.0$\times 10^{-3}$ Jy/beam). The cyan and white star shows the peak intensity in the continuum source position for all the images. The beam size is represented by a yellow ellipse in the bottom left corner.}
\label{second_outflow}
\end{figure*}

There is an elongated and clumpy structure northeast of Par-Lup3-4 with a size of $\sim5^{\prime\prime}$ and a velocity gradient from 5.44 to 7.20 km/s whose origin is unknown (see Fig~\ref{second_outflow}, right panel).  This structure seems to originate close to the position of Par-Lup3-4, with a velocity that increases with distance. It ends in the north at the position of a more extended clump, which may be part of the surrounding parental molecular cloud. We discuss the possible nature of this feature in the next section.

\subsubsection{CO(2-1)}
The first molecular gas emission detection (bottom panel in Fig. \ref{figure:zero_images_b}) of Par-Lup3-4 was at the frequency of CO(2-1) as part of our ALMA Band 6 Lupus 1 and 3 dataset. The CO(2-1) emission spans a total velocity of $\sim$13 km/s in velocity channels ranging from -3.2 to 10.0 km/s (see Appendix \ref{channel_map_CO2-1}), and it shows similar spatial and spectral characteristics as CO(3-2). 
Its spectrum has a double-peak profile, with a more intense red wing and a self-absorption feature between 2.3 to 4.5 km/s (see Fig. \ref{espectro_central}).  The CO(3-2) emission described in the previous subsection suggests the existence of a compact bipolar outflow, which is confirmed by the CO(2-1) emission line detected in our Band 6 data. 

The blueshifted emission spans velocities between -3.2 to 2.8 km/s with two spatial components that show an arc-like structure in the southeast direction (from -3.2 to 1.5 km/s) and in the northwest (from 1.8 to 2.5 km/s). The redshifted emission (velocities between 3.7 to 10 km/s) shows a similar trend, tracing an arc-like shape in the northwest direction between $\sim$4 to 6.6 km/s and in the southeast direction between 7.0 to 10 km/s (see Appendix \ref{channel_map_CO2-1}). These blue- and redshifted structures for the first time suggest that a compact low-velocity bipolar molecular outflow near the plane of the sky is powered by Par-Lup3-4, with the southern lobe showing higher velocities than the systemic velocity than does the northern lobe. Integrated red- and blueshifted CO(2-1) contour maps are provided in Appendix \ref{red_blue_b6}.

Extended emission and negative features due to the effects of filtering large-scale structures by the interferometer are present near 2.7, 4, and 5 km/s. In particular, the remnants of the parental cloud are seen in velocity channels from 3.4 to 4.3 km/s, and this might be a remnant of the envelope in which the source was originally embedded.

The northern stream of clumpy material (hereafter called possible secondary outflow) observed in CO(3-2) at velocities between 5.6 to 7.2 km/s is also detected in the CO(2-1) transition, with similar characteristics in terms of speed and location. The nature of this structure is discussed in Sect.\ref{sec:Molecular_outflow}.

\subsubsection{$^{13}$CO(3-2)} \label{13co}
$^{13}$CO(3-2) emission is detected very close to Par-Lup3-4 in a velocity range between  0.96 and 7.12 km/s (see Appendix \ref{channel_map_13CO}).  $^{13}$CO(3-2) traces a more compact and denser structure than the structure that is observed with the other two $^{12}$CO molecular transitions, where the northwestern outflow cavity is still well perceived (top right panel in Fig. \ref{figure:zero_images_b}).   

\begin{figure}[ht]
\includegraphics[width=0.5\textwidth]{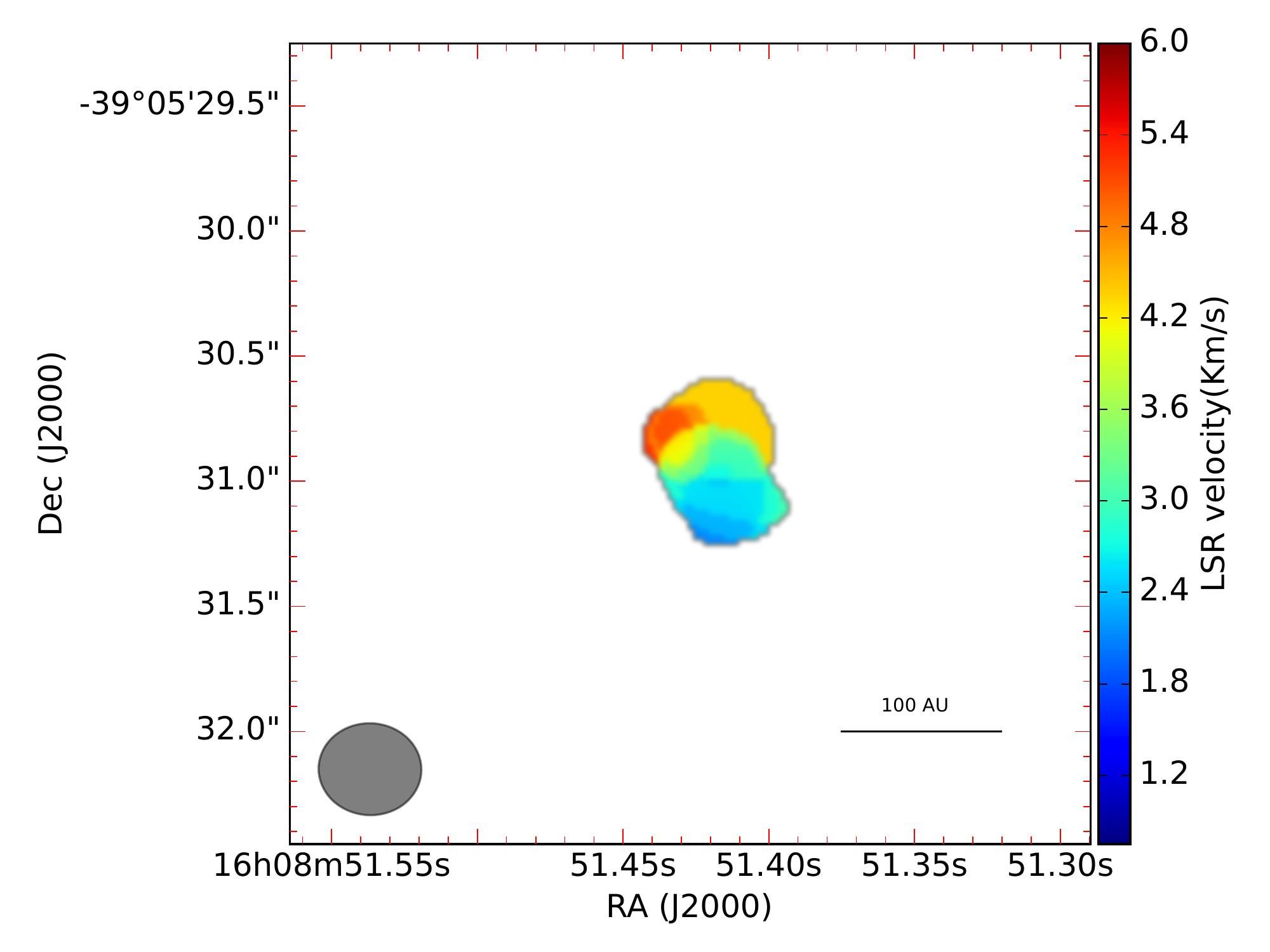}
      \caption{Velocity map of the $^{13}$CO emission line with a threshold of seven times the rms (4.07 $\times 10^{-3}$ Jy/beam). The beam size is represented by the gray ellipse in the bottom left corner.}
\label{espectro_central_13C0}
\end{figure}

\begin{figure}
\centering
\includegraphics[width=\hsize]{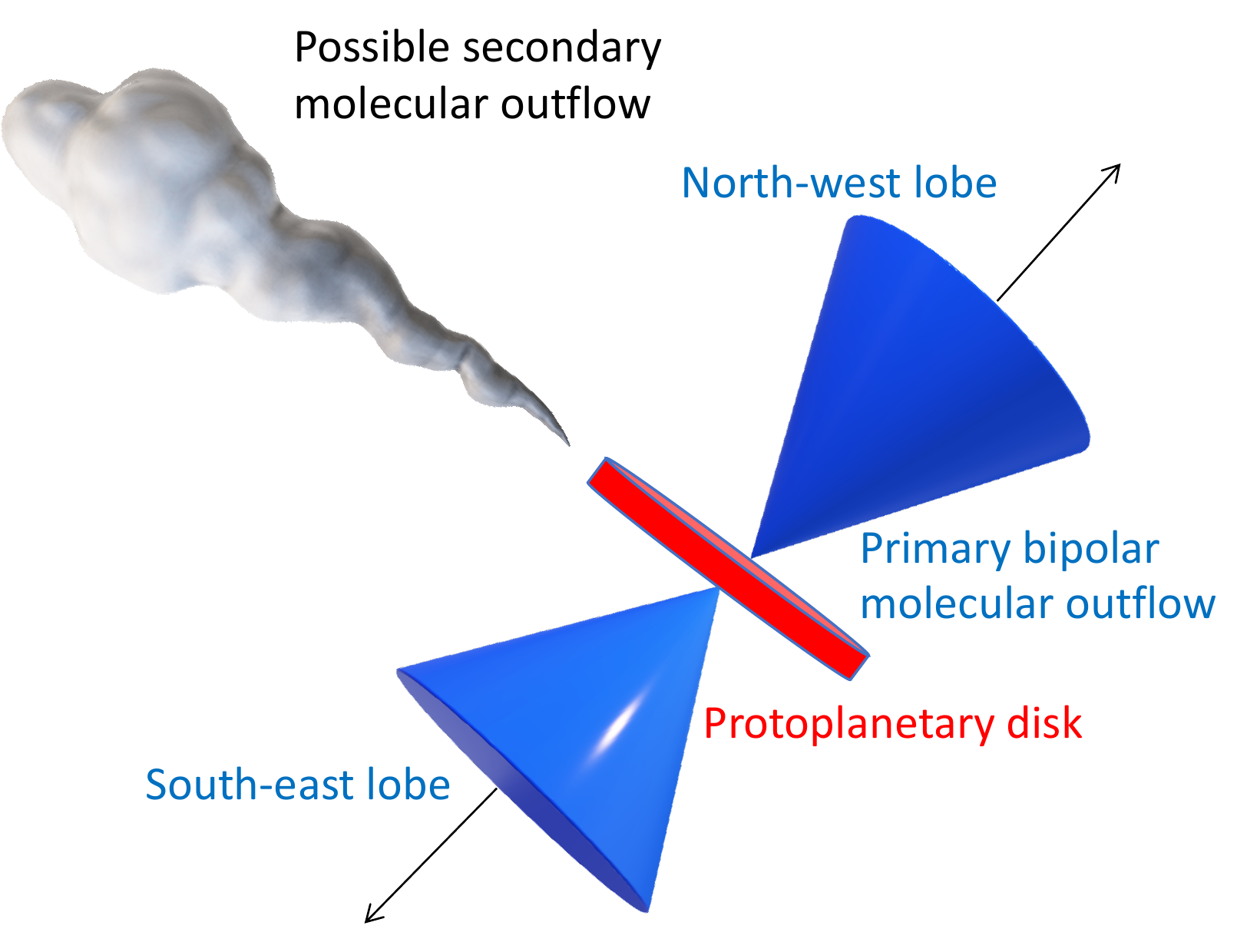}
      \caption{Schematic picture of the relative position of the primary bipolar molecular outflow, the protoplanetary disk, and the possible secondary molecular outflow.}
\label{esquema}
\end{figure}

To investigate the nature of the $^{13}$CO line, we calculated the optical depth and obtained a value of 0.31 that is in the optically thin regime. The $^{13}$CO spectrum is less affected by possible cloud contamination (compared to the $^{12}$CO spectra),  and we used it to infer a more accurate value for the systemic velocity of the source, that is, the velocity average of gas, which we found to be between 3.46 to 3.90 km/s. The velocity map (Fig. \ref{espectro_central_13C0}) of the most intense and compact gas close to Par-Lup3-4 suggests a rotation pattern with redshifted material in the northeast and blueshifted material in the southwest, in a flattened structure perpendicular to the direction of the molecular outflow that is clearly detected in the other CO transitions. Moreover, the velocity pattern is oriented in the expected direction of the circumstellar disk. The spatial and spectral resolution as well as the sensitivity of our data do not allow us to confirm the Keplerian nature of this rotation. Future ALMA observations are required to confirm and study this structure properly. 
A schematic view of the relative position of the protoplanetary disk, the primary bipolar molecular outflow, and the possible secondary outflow are in Fig. \ref{esquema}.

\subsection{Geometrical and dynamical properties of the outflow}
We derived the geometrical and dynamical properties of the primary outflow of Par-Lup3-4 to add evidence for the characterization of outflows surrounding VLM stars and BDs. The calculated geometrical properties include the opening angle, the average length, and the position angle of the outflow, which were measured by hand for both lobes. Our ALMA observations are consistent with the base of a molecular outflow (section \ref{sec:co_3-2_descripcion}), but neither the full ellipsoidal lobe nor the shockwave front are detected. Therefore we can only obtain a lower limit for the average lobe length. The average length was measured using the 3$\sigma$ contour of the CO(3-2) flux-integrated map (Fig. \ref{jet}), and we obtained values of 2.5 arcsec for the southeast lobe and 1.1 arcsec for the northwest lobe. The average size is 1.8 arcsec. The opening angle was calculated in this map, providing an angle of 108$^{\circ}$ for the southern lobe and 116$^{\circ}$ for the northern lobe. Additionally, we measured a position angle of 116$^{\circ}$ and 307$^{\circ}$ for the southern and northern lobes, respectively.

The dynamical properties we studied comprise the outflow mass (M$\mathrm{_{outflow}}$), the dynamical time ($\tau\mathrm{_{dyn}}$), the momentum (P$\mathrm{_{outflow}}$), the kinetic energy (E$\mathrm{_{outflow}}$), the luminosity (L$\mathrm{_{outflow}}$) and the force (F$\mathrm{_{outflow}}$). The brightness temperature of the CO(3-2) line peak (20.6 K) is similar to the kinetic temperature, which indicates that at systemic velocities, the line is optically thick ($\tau$\,>>\,1).  We therefore derived an excitation temperature (T$\mathrm{_{ex}}$) of $\sim$26.4 K from this spectrum. This value is similar to the value that was obtained in previous studies of similar sources \citep{Phan-Bao08-1, Phan-Bao11-1, Phan-Bao14-1}, where T$\mathrm{_{ex}}$ ranges between 20 and 35 K.  

\vspace{7mm}
The column density and the mass of the primary CO outflow were calculated following the prescription in \citet{Scoville1986} and \citet{Palau2007} for the CO(3-2) transition.  The optical depth, measured channel by channel in the four arc-like outflow structures described in section \ref{sec:co_3-2_descripcion},  provides values of $\tau$ <<1.  We obtained a mean optical depth value of 0.25 for both the blueshifted and redshifted southwest cavity as well as for the redshifted northwest cavity, and a value of 0.15 for the blueshifted northwest cavity. Therefore we consider that the wings of the CO(3-2) emission line are in the optically thin regime. Emission close to the position of the central object is optically thick, and therefore this emission was excluded from calculations. Consequently, our estimates of the outflow mass are lower limits because the border between the central object and the base of the outflow is almost indistinguishable. 

We obtained a total outflow mass of 9.5$\times$10$^{-7}$ M$_{\odot}$ as the sum of the mass of the four arc-like structures: the sum of 3.4$\times$10$^{-7}$ M$_{\odot}$ and 2.3$\times$10$^{-7}$ M$_{\odot}$ for the blueshifted ([-1.60 km/s,2.36 km/s]) and redshifted ([6.76 km/s, 9.84 km/s]) components of the east lobes plus the sum of 9.8$\times$10$^{-8}$ M$_{\odot}$ and 2.8$\times$10$^{-7}$ M$_{\odot}$ for the blueshifted ([1.48 km/s, 2.36 km/s]) and redshifted ([4.12 km/s, 5.88 km/s]) components in the west lobes.   
The outflow velocities extend to -2.7 km/s in the blueshifted lobe  and to 9.8 km/s  in the redshifted lobe, resulting in a v$\mathrm{_{max}}$ of $\sim6$\ km/s.  In our calculations we did not apply an outflow inclination correction because the brighter emission of the CO(3-2) at the borders of the cavity suggest a limb-brightening effect, and therefore also that this emission is mostly dominated by material closer to the plane of the sky.

The remaining dynamical parameters were obtained using the formulas in Table \ref{tab:formulas}. Their values are $\tau\mathrm{_{dyn}}$ = 220 yr, P$\mathrm{_{outflow}}$ = 5.7$\times$10$^{-7}$ M$_{\odot}$ km/s,  E$\mathrm{_{outflow}}$= 3.4$\times$10$^{38}$ erg, L$\mathrm{_{outflow}}$= 1.2$\times$10$^{-5}$ L$_{\odot}$ , and F$\mathrm{_{outflow}}$ = 2.6$\times$10$^{-8}$ M$_{\odot}$ km/s yr. 

We finally warn that the values obtained in this section should be treated with
some caution.  Although these measurements are useful for constraining models, it is highly probable that not the entire outflow structure is detected, and molecular cloud contamination affects several emission channels near the cloud velocity. We estimate that other uncertainties such as T$\mathrm{_{ex}}$, optical depth, and the outflow extent may result in an increase in outflow mass by less than a factor 2.

\begin{table}[!hbt]
\caption{Geometrical and dynamical properties of the outflow. Because the flux mass is probably underestimated (see \ref{maximum}), these values should be regarded as order-of-magnitude estimates.}             
\label{tab:formulas}
\resizebox{0.5\textwidth}{!}{
        \begin{tabular}{l l l}        
            \hline\hline                 
            Derived  & Formula & Value \\
            properties & \\
            \hline\hline 
            Length &  R$\mathrm{_{lobe}}$  & 1.8 arcsec\\
            Velocity &   v$\mathrm{_{max}}$ & 6 km/s \\
            Dynamical time & $\tau\mathrm{_{dynamical}}$ =  R$\mathrm{_{lobe}}$  /  v$\mathrm{_{max}}$ & 220 yr \\ 
            Mass-loss rate & $\dot{\mathrm{M}}$ = M$\mathrm{_{outflow}}$ / $\tau\mathrm{_{dynamical}}$ & 4.3$\times$10$^{-9}$ M$_{\odot}$/yr \\
            Momentum & P$\mathrm{_{outflow}}$ = M$\mathrm{_{outflow}}$  $\times$ v & 5.7$\times$10$^{-7}$ M$_{\odot}$ km/s \\
            Energy & E$\mathrm{_{outflow}}$ = 1/2 M$\mathrm{_{outflow}}$  $\times$ v$^{2}$ &  3.4$\times$10$^{38}$ erg\\
            Luminosity & L$\mathrm{_{outflow}}$ = E$\mathrm{_{outflow}}$ /$\tau\mathrm{_{dynamical}}$ & 1.2$\times$10$^{-5}$ L$_{\odot}$\\
            Force & F$\mathrm{_{outflow}}$ = P$\mathrm{_{outflow}}$ / $\tau\mathrm{_{dynamical}}$ & 2.6$\times$10$^{-8}$ M$_{\odot}$ km/s yr \\
        \end{tabular}
}        
\end{table} 

\label{gaseo}

\subsection{Spectral index}
The spectral index ($\alpha$) at millimeter frequencies used to be a tool for inferring grain growth signatures under the assumptions that the emission is in the Rayleigh-Jeans regime and is optically thin. The spectral index in the optically thin regime provides information on the dust opacity index ($\beta$); its value depends on dust size, and consequently, on grain growth in the case of large dust particles. Typical $\alpha$ values for the interstellar medium (ISM) are close to 3.7 \citep{Draine06}, while lower values $\leq$ 3 are found in most disks, and they are interpreted as a signature of grain growth \citep{Ricci10-1, Ricci10-2, Ribas17-1}. We assume that the emission is associated with the Rayleigh-Jeans regime and that the optically thin  emission $\alpha$ depends on the flux as F$_{\nu} \propto \nu^{\alpha}$, where  $\beta$ = $\alpha$ -2 and $\kappa_{\nu} \propto \nu^{\beta}$. 

We obtained the spectral index, $\alpha$, using the formula
\begin{equation}
\alpha = \frac{\mathrm{log\frac{F_{\nu_{1}}}{F_{\nu_{2}}}}}{\mathrm{log}\frac{\nu_{1}}{\nu_{2}}}
,\end{equation}
where F is the flux density at frequencies $\nu_{1}= 328$ GHz and $\nu_{2}= 225$ GHz. The  $\alpha$ uncertainty is determined as  
\begin{equation}
\Delta \alpha =\sqrt{\mathrm{\bigg(\frac{\Delta S_{1}}{S_{1}ln\big(\frac{\nu_{1}}{\nu_{2}}\big)}\bigg)^{2}}+\mathrm{\bigg(\frac{\Delta S_{2}}{S_{2}ln\big(\frac{\nu_{1}}{\nu_{2}}\big)}\bigg)^{2}}}
.\end{equation}
Using the values  given in Table \ref{tabla_ppal}, we obtained that the Par-Lup3-4 spectral index is 1.6 $\pm$ 0.4. $\alpha$ values below 2 imply that the disk does not appear to follow the Rayleigh-Jeans approximation  \citep{Ansdell18-1}. Therefore we cannot obtain information about the composition and size of the dust using the spectral index for Par-Lup3-4.

\section{Discussion}
\label{discusion}

\subsection{Nature of the extended gas emission to the northeast}
\begin{figure}
\includegraphics[width=0.5\textwidth]{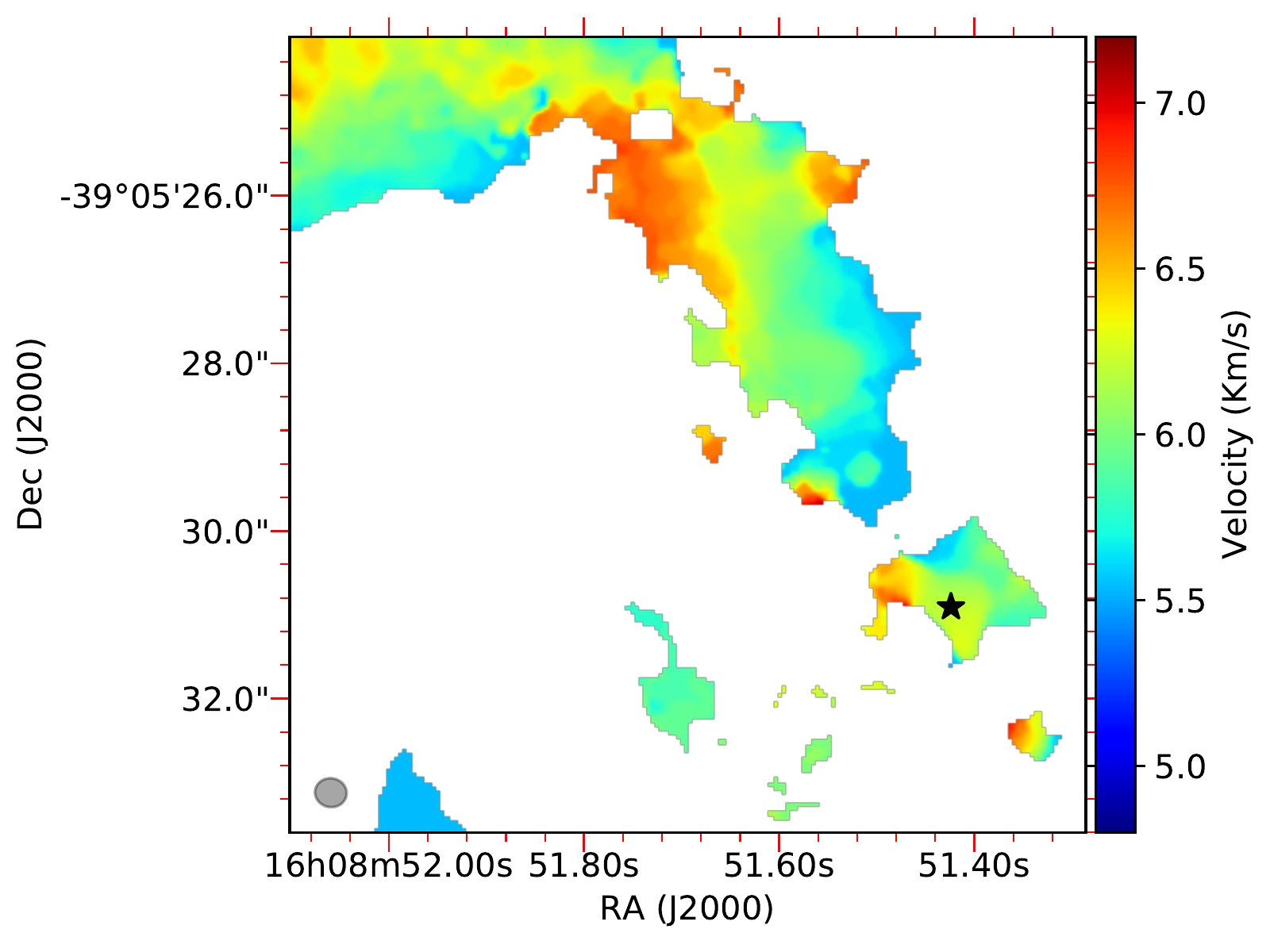}
\includegraphics[width=0.5\textwidth]{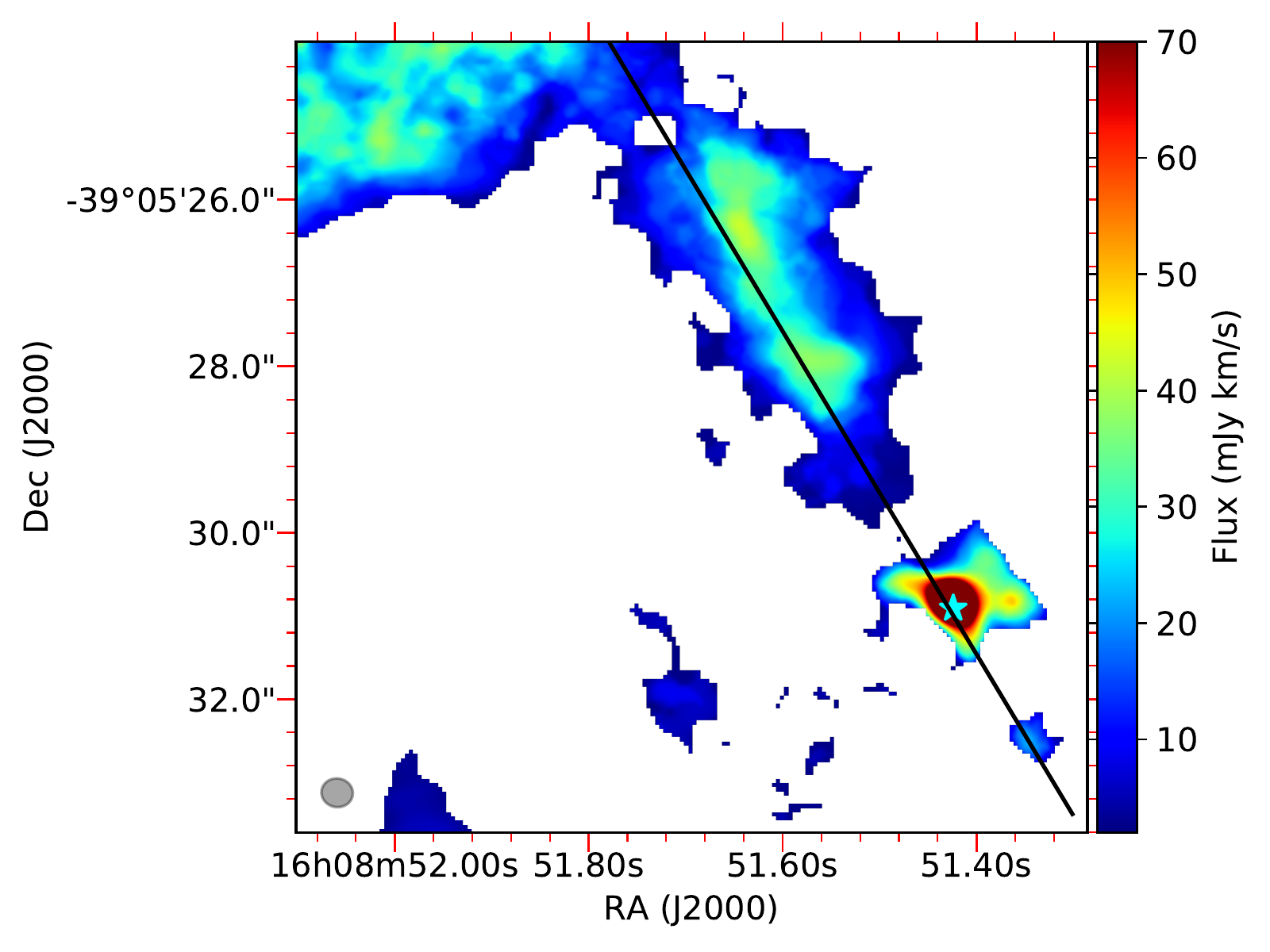}
\centering
\includegraphics[width=0.5\textwidth]{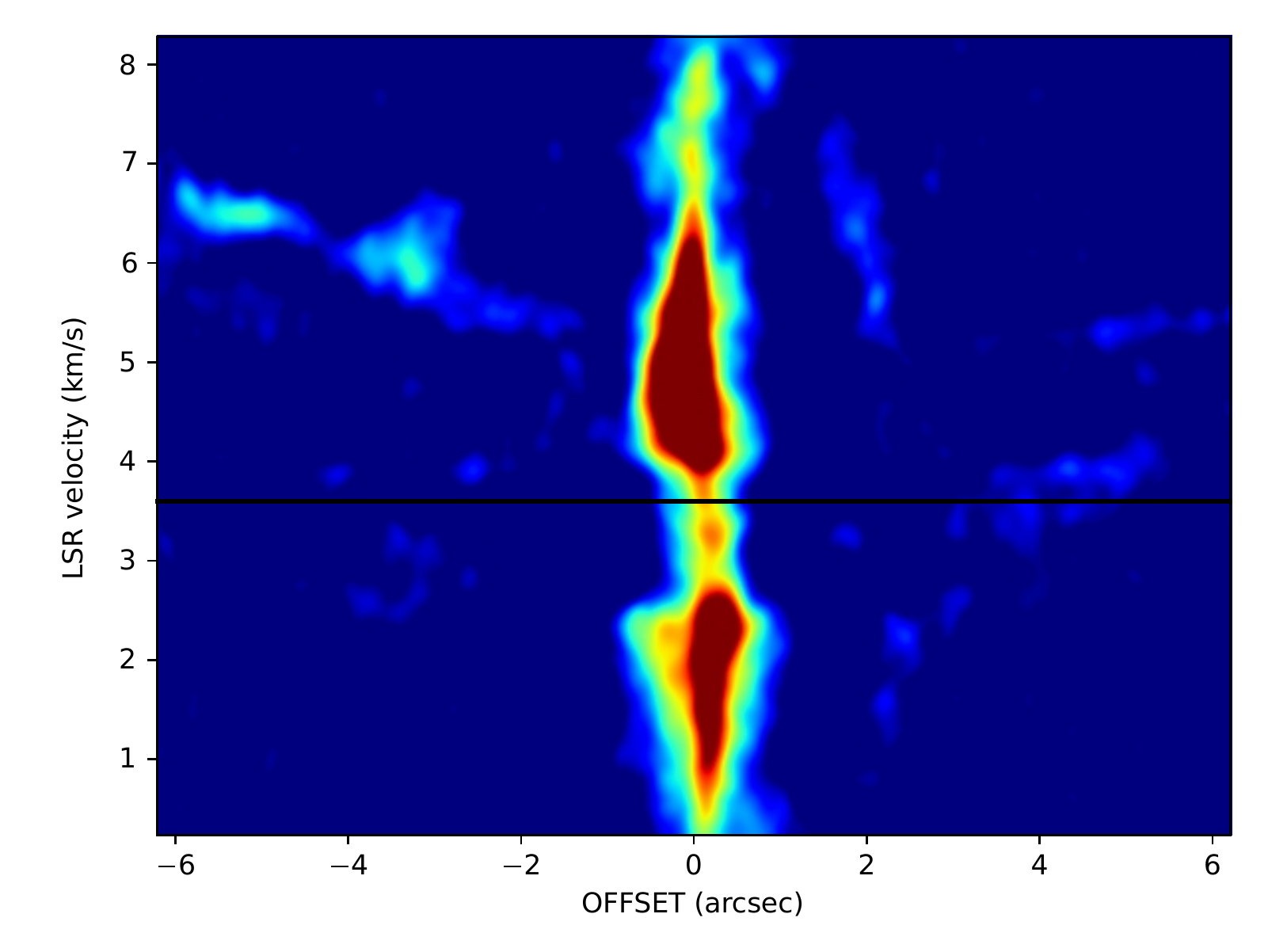}
\caption{Images of the CO(3-2). Top panel: ALMA velocity  map considering an intensity above 5$\sigma$. Middle panel: CO(3-2) flux-integrated ALMA map with pixels above 5 $\sigma$. The black line marks the path for the position-velocity diagram in the bottom panel. Bottom panel: Position-velocity diagram with a PA of 30.5 deg and a width of  $\sim$1.2 arcsec. The horizontal black lines shows the systemic velocity of Par-Lup3-4.  The cyan or black star marks the position of the peak intensity in the continuum image.
In the two first images the beam size is represented by the gray ellipse in the bottom left corner.} \label{fig:CO_momentos_0_1_outflow}
\end{figure}

This subsection discusses the nature of the elongated and clumpy structure observed in CO(3-2) (Fig. \ref{fig:CO_momentos_0_1_outflow}) and CO(2-1). The structure is redshifted with respect to the LSR of Par-Lup3-4 and propagating from the center toward the northeast side of the field of view, with less redshifted velocities toward the position of the source. Here we discuss four possible origins of this structure.

First, it is possible that this structure is a second molecular outflow originating from Par-Lup3-4. This could be explained if Par-Lup3-4 is not a single object but a close binary system. Adaptive optics (AO)-assisted observations \citep{Huelamo10-1} have not revealed an additional companion down to 0.1 arcsec ($\sim$15.5 pc), but closer companions cannot be excluded. Additionally, another argument in favor of this possibility is that the emission appears to originate from the location of Par-Lup3-4. The emission is detected with a velocity of $\sim$5.4 km/s at a distance of $\sim$2 \ arcsec (310 au) northeast of the source; and up to $\sim$6.7 km/s at a distance of $\sim$6 \ arcsec (930 au) from the source, as is clearly seen in the position-velocity diagram of Fig.\,\ref{fig:CO_momentos_0_1_outflow}. Bands 6 and 7 show very similar morphologies and velocity patterns. If the emission arises from an actual outflow, it appears to be a monopolar one, as there is no counterpart with emission moving toward the southwest. Monopolar outflows have been observed in the literature for low-mass stars \citep{Codella14-1, Louvet18-1}. On a related note, the jet detected by \citet{FC2005} in Par-Lup3-4 shows an asymmetry, as the southeast lobe is much more prominent than the northwest lobe. Although the orientation of this second molecular outflow, perpendicular to the optical jet, is less likely, a similar scenario with two perpendicular outflows has previously been seen in other sources \citep{Tobin15-1}; this might be interpreted as another indication of the binary nature of the central source. 

A second possibility is that we are observing cloud contamination. In support of this hypothesis, we note that the velocity of the structure is very close to  the velocity of the second component of the cloud. In the same low-velocity regime, we also detect extended cloud emission compatible with the cloud velocity. However, we detect very compact structures when we perform a visibility range (see Appendix \ref{fig:CO_momentos_0_1_outflow_filtrado}), and inhomogeneities with this morphology in molecular clouds, although possible, are not common.  It is also not expected that we see clear strips in the field of view pointing toward Par-Lup3-4 along several channels instead of randomly distributed material, as we can see in the channels with a velocity higher than 5.70 km/s in Appendix \ref{channel_map_nube}. 

A third scenario is that the outflow comes from another source that is neither Par-Lup3-4 nor a close companion. This may be the unlikely case of a source in the line of sight of Par-Lup3-4 that would be responsible for the molecular outflow. Another possibility is a nearby source that just happens to be outside the field of view to one edge or another. It might even be a poorly known source in Lupus located at several dozens or hundreds of arcsec away.

Another possibility to discuss is a stream of envelope material that is infalling onto Par-Lup3-4. In case of pure infall of foreground material (i.e., material located between the star and us), we would expect a velocity pattern with more redshifted velocities closer to the central source and we see the opposite, where the pattern is more consistent with outflowing material from a Hubble-type outflow in which the velocity increases with distance from the protostar. Now, if the infalling stream is also rotating clockwise, this material might be slowed down closer to the protostar, providing a velocity pattern more compatible with the observed pattern. 

\label{sec:Molecular_outflow}

The current (sub)mm and optical observations do not allow us to distiguish between the four scenarios. Future studies using ALMA should be able to unveil the origin of the moving structure. ALMA  observations with better velocity resolution and higher sensitivity may also help to distinguish between outflow or cloud emission in the second scenario. High angular resolution observations as well as radial velocity observations in the infrared  may help to  detect a close binary.

\begin{table*}
\centering
\caption{Best-fit parameters from SED modeling.}             
\label{tab:modeling_results}  
\begin{tabular}{l c c}        
\hline\hline                 
Parameter & JHK from 2MASS &  JHK from \citet{Comeron03-1}  \\
\hline
Stellar radius ($R_\odot$)& 2.0 & 2.0 \\
Disk dust mass ($M_\odot$)& 5$\times10^{-7}$ & 1$\times10^{-7}$ \\
Maximum grain size (mm) & 10 &  0.5 \\
Scale height at 100\,au (au)& 20 & 20 \\
Flaring index & 1.2 & 1.1 \\
Surface density index & -1.5 & -1.0 \\
Inclination (deg) & 82.5 & 85 \\
Interstellar extinction (mag) & 3.5 & 2.5 \\
\hline\hline 
\end{tabular}
\end{table*} 

\subsection{SED fitting}
\label{fiteo}
\begin{figure}
\centering
\includegraphics[width=\hsize]{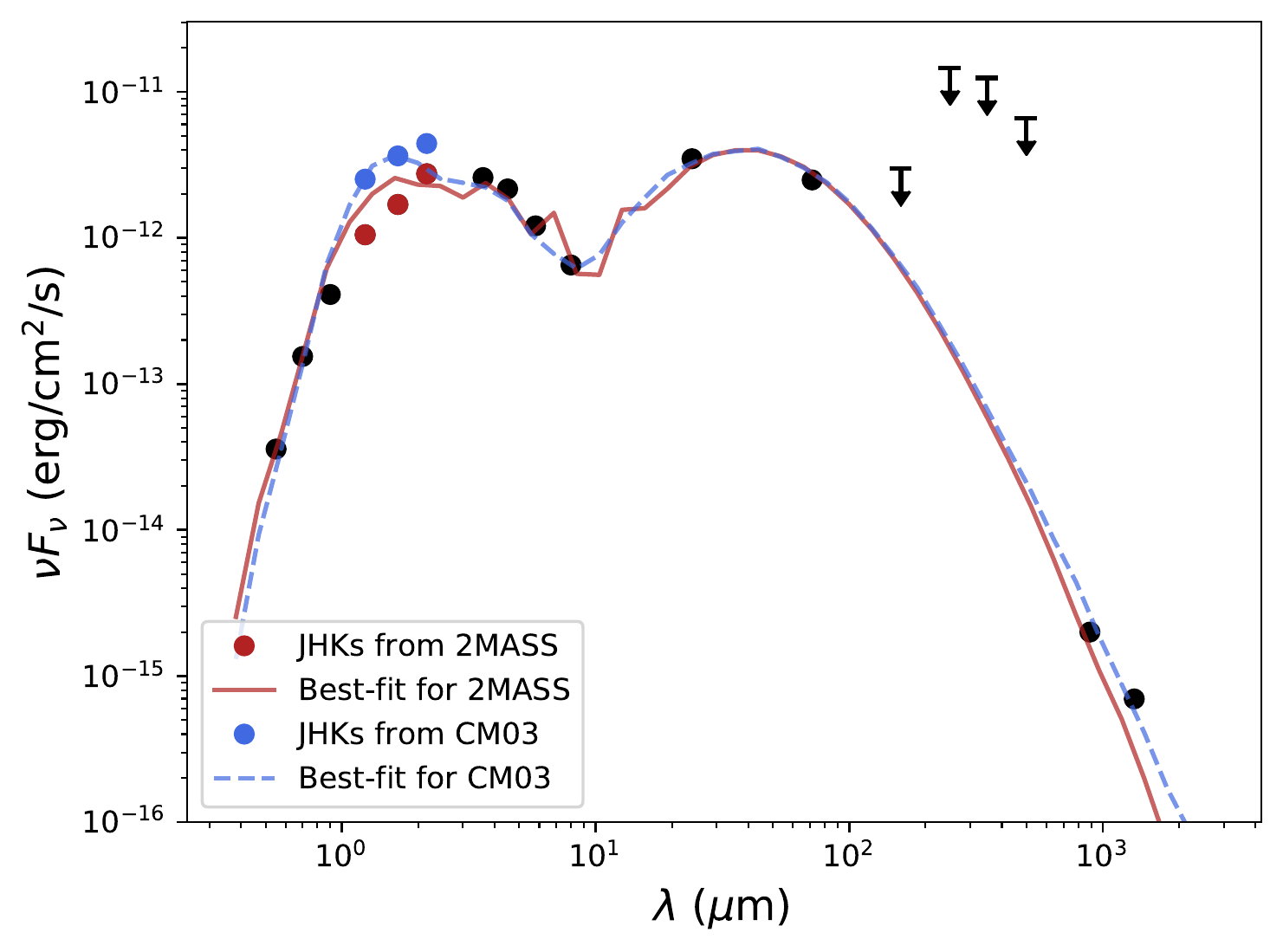}
      \caption[SED and modeling results for Par-Lup3-4.]{SED and modeling results for Par-Lup3-4. Dots are photometric observations. Arrows are upper limits. Red and blue dots are the 2MASS and the \citet[][CM03]{Comeron03-1} observations, respectively. The best-fit models are also shown with a similar color code.}
\label{fig:SED_modeling}
\end{figure}

As part of a study of Par-Lup3-4, we have also tried to characterize its circumstellar disk, taking advantage of the fact that new photometric data are available and can help to populate the SED studied by \citet{Huelamo10-1}. We therefore complemented the photometric points from \citet{Huelamo10-1} with new Herschel (PACS $\text{and}$ SPIRE; see Tab. \ref{tab:fotometria}) and ALMA data. The resulting SED is displayed in Fig. \ref{fig:SED_modeling}. Then, we used the radiative transfer code MCFOST \citep{Pinte2006,Pinte2009} to infer the main disk properties.

\begin{table}
\centering
\caption{PACS and SPIRE photometry}             
\label{tab:fotometria}  
\begin{tabular}{l c c}        
\hline\hline                 
Wavelength & Flux &  Flux error  \\
$[\mu$m] & [mJy] & [mJy] \\
\hline
71.42 & 59.4 &   2.4 \\
160  &     $<$159.2 & -  \\
250  &      $<$1215 &- \\
350  &     $<$1455 & - \\
500  &     $<$1101 &-  \\
\hline\hline 
\end{tabular}
\end{table} 

Modeling protoplanetary disks involves defining several free parameters, some of which are highly uncertain or degenerate. Such a scenario is better dealt with within a Bayesian framework and using statistical tools such as Markov chain Monte Carlo methods, but this approach is computationally very demanding and has only been applied in a few cases \citep[e.g.,][]{Ribas2016, Wolff2017}. Because the amount of data available at long wavelengths for Par-Lup3-4 is limited, we chose to run a grid of models to obtain a general idea of the system parameters.  

Our initial attempts at fitting the SED of Par-Lup3-4 used fixed stellar parameters from the BT-Settl models \citep{Allard2012, Baraffe15-1} based on previous studies. The stellar temperature was fixed to 3200\,K \citep{Alcala17-1} because it is firmly given, with a narrow uncertainty margin, by the spectral classification of the star. Age estimates for the source range from 1 to 3\,Myr \citep{Comeron03-1,Alcala17-1}. Assuming 2\,Myr and using the BT-Settl models, we derived a stellar radius and a mass of 1.1\,R$_\odot$ and 0.2\,M$_\odot$. The distance to the source was set to d$\sim$155\,pc (see Section \ref{info}). Regarding the disk parameters, we have fixed the disk inner radius and the minimum grain size to 0.05\,au and 0.005\,$\mu$m, following \citet{Huelamo10-1}.

For our model, we defined eight free parameters and explored them within reasonable ranges:

\begin{itemize}
    \item the stellar radius, including the value of  1.1\,$R_\odot$ derived from isochrones, plus 1.5, 2, and 2.5\,$R_\odot$ as additional values;
    
    \item the disk dust mass, from $1\times10^{-7}$ to $1\times10^{-5}$\,M$_\odot$ in steps of 0.5 dex;
    
    \item the maximum grain size, from 500\,$\mu$m to 1\,cm in steps of 0.5 dex;
    
    \item the scale height at a radius of 100\,au, from 5 to 20\,au in steps of 5\,au;
    
    \item the flaring index, with values 1.0, 1.1, and 1.2 (the disk is assumed to flare following $H(r)\propto r^\gamma$, where $H(r)$ is the scale height as a function of radius, and $\gamma$ is the flaring index);

    \item the surface density index, with values -0.5, -1.0, and -1.5;
    
    \item the inclination, with values 80, 82.5, and 86\,deg,  based on the inclination derived from the optical jet \citep{Comeron2011-1};
and 
    \item the interstellar extinction values, from 0 to 5\,mag in steps of 0.5\,mag.
    
\end{itemize}

This setup resulted in a significant underestimation of the optical fluxes, even assuming no extinction. \citet{Huelamo10-1} found a similar problem in their modeling efforts of this source, but attributed it to an uncertain distance value. However, the distance measurement and uncertainty in the distance estimate by \emph{Gaia} clearly show that this is not the case, and the source must be intrinsically more luminous in order to reproduce the observed fluxes. For this reason, we then included the stellar radius as a parameter in the subsequent modeling process. 

$\chi^2$ values were then computed for each model. We note that there are two different results from near-IR observations \citep[one from 2MASS and from][]{Comeron03-1}, possibly reflecting the variable nature of Par-Lup3-4. Thus, two different $\chi^2$ values were computed for each model.

The results from the SED modeling are shown in Table~\ref{tab:modeling_results} and Fig.~\ref{fig:SED_modeling}. SED fitting is, in general, a degenerate process, and in our case, the photometric coverage from the IR to mm wavelength is rather poor.  The derived values are highly uncertain (especially when we consider that Par-Lup3-4 may still be surrounded by a dusty envelope, which we did not consider in our models). However, one crucial result is the fact that a radius of 2\,$R_\odot$ is required in both cases. Different tests with varying source distance and radius showed that it is not possible to fit the optical part of the observed SED with a source of 3200\,K and a radius of 1.1\,$R_\odot$ at the distance of 155\,pc measured by \emph{Gaia} because the source is simply not bright enough. Given the small uncertainties associated with the parallax in the \emph{Gaia} DR2 catalog and because its distance is compatible with that of other sources in the region, a plausible explanation is that the radius of Par-Lup3-4 is larger, and thus it is younger than the median age of Lupus. The radius is even inconsistent with models at any age. This idea is also supported by the presence of a molecular outflow, and by the fact that the source appears to be still embedded in the cloud. If this is the case, it is likely that Par-Lup3-4 is still surrounded by material from the envelope. 

A second possibility to explain the 2 R$_{\odot}$ radius, and therefore the higher luminosity of Par-Lup3-4, is that the source is a binary system with a mass ratio near one. If this is the case, the perpendicular structure described in section\,\ref{sec:Molecular_outflow} could be an outflow originating from the companion. However, as explained before, no clear signature of a companion has been found so far. If any of these two scenarios applies, the disk parameters derived from our modeling process should be treated with caution because the models we used may not reflect the true nature of the source.

We note that there is a factor of $\sim$2 to $\sim$27 in difference between the dust disk mass obtained from fitting the SED and the values obtained in Section \ref{resultados:contin}. We acknowledge that the model itself may need to be refined in order to minimize this discrepancy.

The SED fitting reported in \citet{Huelamo10-1} fixed the stellar parameters from BT-Settl models, but they found that it was not possible to fit the SED unless they changed the distance.  Now, using Gaia, we are certain that the distance is not the main problem. While it is impossible fit the SED by assuming a single central object of a plausible luminosity, the existence of a secondary outflow leads to the reasonable explanation that Par-Lup3-4 is formed by a close binary with two equal-mass components.

\subsection{Characterizing the molecular outflow cavity}
Par-Lup3-4 is the first VLM star to date for which we detected the base of a bipolar molecular outflow at (sub)mm wavelengths, and for which the highly supersonic outflow (jet) has been detected at optical wavelengths. Following the low-mass star outflow model, the interaction between the jet or the wide-angle outflow and the envelope creates the detected cavities \citep{Li-1996}. The expelled gas and material carves out the cavities in the envelope, and the interaction in the boundary between the outflowing gas and the envelope material creates the physical conditions for exciting the CO transitions that we detected. In this section, we characterize the VLM outflow of Par-Lup3-4 in the context of mass ejection from low-mass protostars. 

The first property that we use to characterize the outflow is the lobe length, which in the case of the primary outflow of Par-Lup3-4 is  $\sim$1.8 arcsec ($\sim$295 au). This  is a lower limit. The uncertainty on the lobe length is at least about one half of the beam size at a distance of $\sim$155 pc, which yields a value of $\sim$28 au, and it is likely much more as we cannot recover the entire lobe. The length of this outflow is on the same order of magnitude as the lengths of other VLM protostars and proto-BDs within the uncertainties, such as ISO-Oph\,102, GM\,Tau, MHO\,5 \citep{Phan-Bao14-1}, IC348-SMM2E \citep{Palau14-1} , or L1148-IRS \citep{Kauffmann11-1}, which have outflows between 500 to 1800 au. The outflow length for low-mass stars is between 0.1 to 10 pc (20,000 and 2,000,000 au) \citep[and references therein]{Arce-2007}, which is one or two orders of magnitude larger than the sizes of outflows from VLM protostars and proto-BDs.  Therefore, the length of the Par-Lup 3-4 outflow, comparable to those of VLM protostars and proto-BDs, is a scaled-down version of outflows from low-mass protostars. 

The outflow velocity measured for Par-Lup3-4 is slightly higher (6 km/s) but on the same order as the velocities observed in other VLM stars and BDs, which are between 1 to 4.7\,km/s \citep{Phan-Bao08-1, Kauffmann11-1, Phan-Bao14-1}. Low-mass stars have outflows with velocities in the range between 10-100 km/s \citep[and references therein]{Arce-2007}. The velocity of the Par-Lup3-4 outflow is closer to the VLM regime than that of low-mass stars.

Par-Lup3-4 has a molecular outflow mass of $\sim$10$^{-6}$ M$_{\odot}$ that is in the range of the observed values for other VLM stars and BDs that span 10$^{-4}$ to 10$^{-6}$ M$\odot$ \citep{Phan-Bao14-1}. As pointed out by \citet{Phan-Bao14-1}, the VLM outflow values are at least one order of magnitude lower than the values obtained for low-mass protostars in a similar evolutionary status. 

The mass-loss rate from Par-Lup3-4 ($\mathrm{\dot{M}_{outflow}}$=M$\mathrm{_{outflow}}$/$\tau{\mathrm{_{dyn}}}$) is 4.3$\times$10$^{-9}$ M$_{\odot}$/yr. The outflow mass-loss rate for a typical low-mass protostar ranges from  8.9\,$\times$\,10$^{-9}$ to 10$^{-4}$\,M$_{\odot}$\,yr$^{-1}$ \,M$_{\odot}$ yr $^{-1}$, although the median value is 10$^{-7}$ M$_{\odot}$ yr $^{-1}$ \citep{Levreault1988-2}.   The  mass-loss rate for VLM stars and BDs is lower, with values lying between  2.5\,$\times$\,10$^{-9}$ and 2\,$\times$\,10$^{-7}$ \,M$_{\odot}$\,yr$^{-1}$ \citep{Phan-Bao14-1}. The mass-loss rate for Par-Lup3-4 is even lower than the expected value for VLM stars and BDs, but this might be an effect of the potentially missed flux emission from the whole outflow extent that we do not observe. 

\begin{figure}
\centering
\includegraphics[width=\hsize]{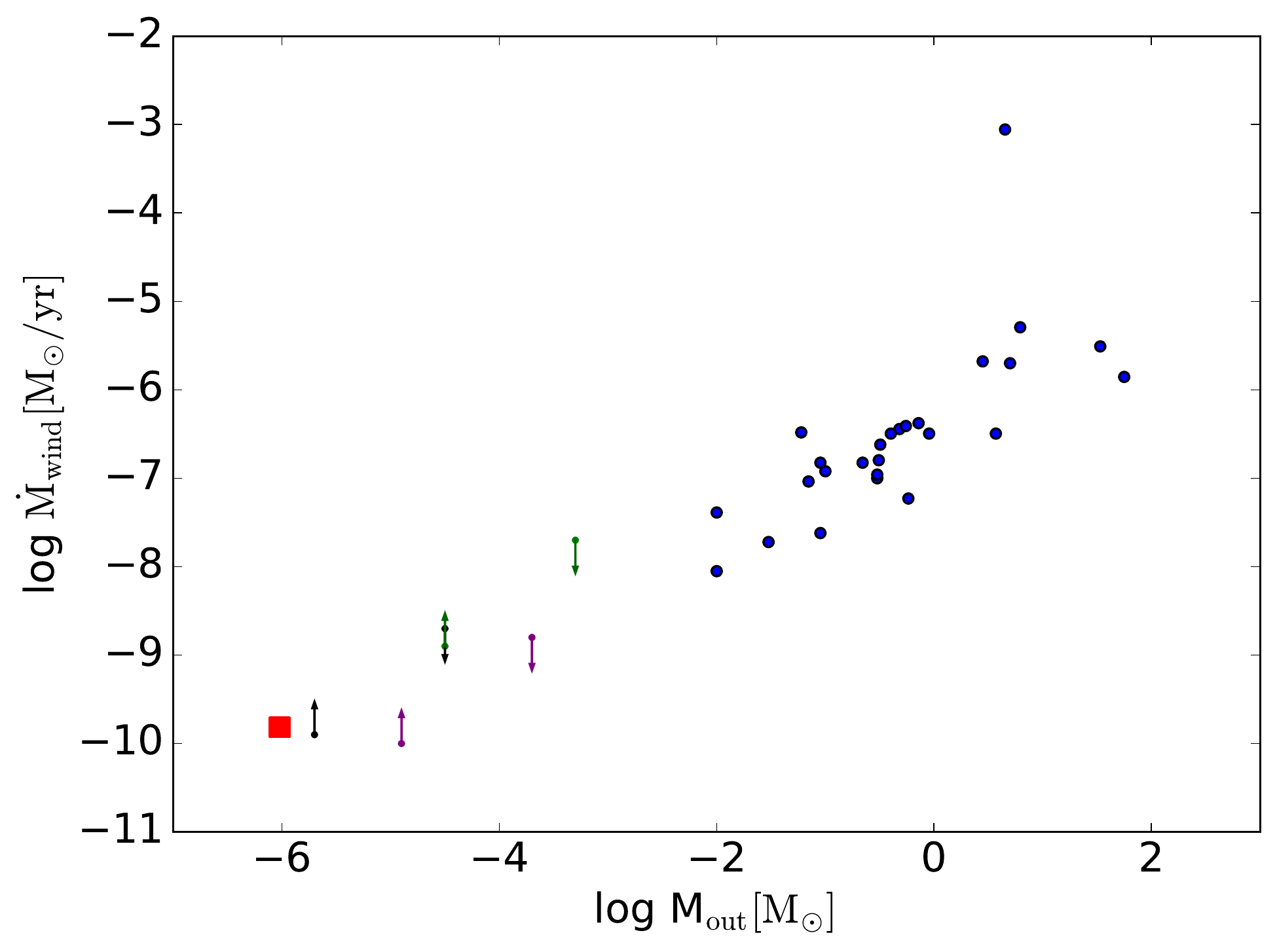}
      \caption[Molecular outflow versus wind mass-loss rate of Par-Lup3-4 and very low-mass sources]{Molecular outflow mass vs. wind mass-loss rate of Par-Lup3-4 (red square), very low-mass sources (arrows are lower and upper limits) from \citet{Phan-Bao14-1}, and Class II young stellar objects (blue circles; \citealt{Levreault88-1}).}
\label{mout_mwin}
\end{figure}

The mass-loss rate of the stellar wind was obtained as $\mathrm{\dot{M}_{wind}}$ [M$_{\odot}$/yr]=M$\mathrm{_{outflow}}$v$\mathrm{_{max}}$/ $\tau{\mathrm{_{dyn}}}$v$\mathrm{_{wind}}$\, \citep{Phan-Bao14-1}. We used a wind velocity of 168 $\pm$ 30 km/s \citep{Comeron2011-1}. We assumed that the momentum from the jet is completely transferred to the molecular outflow, which may occur in Class II sources. The wind mass-loss rate derived for Par-Lup3-4 is 1.53$\times$10$^{-10}$ M$_{\odot}$ /yr, which is very similar to the value of MHO5, another VLM star studied by \citet{Phan-Bao14-1}. We compare the outflow mass against the mass-loss rate of the stellar wind for Par-Lup3-4 along with other VLM stars, BDs, and low-mass stars using low-resolution observations in Fig. \ref{mout_mwin}, and it appears that Par-Lup3-4 follows the trend of more massive sources.

Accretion and outflow, jet, or winds are phenomena that are deeply linked \citep{Hartigan1995, Calvet1997, Rigliaco-2013, Natta14-1}. As the material infalls from the envelope or disk onto the central source, a jet perpendicular to the disk is launched by angular momentum conservation. The accretion rate in combination with the outflow properties can give us information about the history of the sources. The accretion rate of Par-Lup3-4 has been measured in recent years: 1.4\,$\times$\,10$^{-9}$\,M$_{\odot}$\,yr $^{-1}$ \citep{Comeron03-1}, 7.9\,$\times$\,10$^{-10}$ \citep{Bacciotti2011}, 5.0\,$\times$\,10$^{-10}$\,M$_{\odot}$\,yr $^{-1}$ \citep{Whelan14-1}, and 4.3\,$\times$\,10$^{-12}$\,M$_{\odot}$\,yr $^{-1}$ \citep{alcalaetal14-1}.  These studies used different extinction values and accretion tracers, and each study had its own assumptions and caveats. The ratio $\mathrm{\dot{M}_{wind}}$/$\mathrm{\dot{M}_{acc}}$ for VLM stars and BDs is between 0.05 to 100 (see Table 3 in \citealt{Phan-Bao14-1}), while for low-mass stars the expected value is in the range of ${\sim}$0.0003-0.4 \citep{Hartigan1995}. When we use the obtained wind mass-loss rate and the accretion rates from the literature, the relation $\mathrm{\dot{M}_{wind}}$/$\mathrm{\dot{M}_{acc}}$ varies between 0.07 to 22, which is in the expected range for VLM stars and BDs.

Additionally, we investigated another parameter, the opening angle, that can be used for evolutionary classification. The wide opening angle of the outflow, described in section \ref{gaseo}, has an average value of 112$\mathrm{^\circ}$ between the two lobes. Previous studies of low-mass stars have discussed the relation between the age or evolutionary classification and the opening angle because the angle broadens with age \citep{Offner-2011}. \citet{Arce-2006} classified low-mass stars as Class 0 with opening angles $\leq$\,55$\mathrm{^\circ}$, Class\,I if the angle is $\geq$\,75$\mathrm{^\circ}$, and Class II when the outflow has no clear structure. A year later, \citet{Arce-2007} defined the boundary for Class I as $\geq$90$\mathrm{^\circ}$. Par-Lup3-4 is consistent with Class I based on these studies and is near to the transition to Class II given the wide opening angle, although these values are still under discussion in the field and cannot be considered as the only parameter to classify the evolutionary stage of the source. Unfortunately, there are no previous records in the literature about the opening angle in VLM stars and BDs. Future studies should correlate opening angle and evolutionary stage in VLM stars using larger samples. 

\subsection{VLM stars and BDs as a scale down version of low-mass stars?}
In the previous section, we reviewed the outflow properties of Par-Lup3-4. Properties such as the outflow length, the velocity, outflow mass, and wind mass are in the expected range for VLM protostars and proto-BDs. Additionally, other properties such as the disk mass are in the expected range for VLM stars and BDs. The disk mass of Par-Lup3-4 is $\sim$ 10$^{-6}$ M$_{\odot}$ with the assumptions mentioned in Section \ref{resultados:contin}. Low-mass protostar disk masses are in the range of 10$^{-3}$ to 10$^{-1}$ M$_{\odot}$ , and the theoretical values obtained for VLM stars and BD disks using radiative transfer algorithms extends from 10$^{-6}$ to 10$^{-3}$ M$_{\odot}$. Recently, \citet{Sanchis20} measured the mass of several BD disk in Lupus and obtained masses between 7 $\times$ 10$^{-4}$ M$_{\odot}$ and 6 $\times$ 10$^{-5}$ M$_{\odot}$. The disk mass measured for Par-Lup3-4 therefore indicates a downsized version of the low-mass protostar disks, although the mass inferred from our ALMA observations may represent a lower limit given the disk inclination and the uncertainties with the optical depth. All these properties are in agreement with previous studies of VLM stars and BDs (e.g., \citealt{Phan-Bao14-1}), which pointed out that the formation of VLM stars and BDs follows a scaled-down version of low-mass star formation.

While these characteristics indicate a downsized version of star formation, there are still several uncertainties in the measurements that may further constrain this. For example, uncertainties are related to the outflow mass calculation, such as the optical depth or the CO abundance relative to the H$_{2}$, that can vary by a factor of three \citep[and references therein]{Dunham2014}. In the case of Par-Lup3-4, all these uncertainties are negligible when we compare the geometry of the outflow, which is not fully revealed with our interferometric ALMA data. This directly affects the average length of the outflow and the dynamical time, propagating the uncertainties into the dynamical parameters, which may be longer if we are able to reveal the whole extent and shape of both outflow lobes. Another important source of uncertainty is the missing flux and possible faint extended emission in our observations; the maximum outflow length that we detect is 2.9 arcsec, but it is expected to have larger scales because we do not detect the shockwave front. Therefore, our results might be impaired by two effects: extended flux was filtered with the interferometer, and the sensitivity is not high enough. 
\label{maximum}

Additionally, previous studies of VLM stars and BDs are biased by low-sensitivity observations, showing a smaller velocity range. This affects the dynamical properties.  Another important bias in previous studies comes from the inclusion of face-on disks or disks with inclinations up to $\sim$30$\mathrm{^{\circ}}$ , which are easier to detect than edge-on disk systems. Previous observation of molecular outflows in VLM stars and BDs might be affected by the limited sensitivity that affects the low detection rate found so far (e.g., \citealt{Phan-Bao14-1}). With this work, we proved that we can detect a very low amount of expelled mass with the high ALMA sensitivity with enough resolution to observe the base of the outflow. Therefore we conclude that more sensitive observations are indispensable to improve the statistics of outflows in these sources. In spite of the uncertainties and the small and biased sample of VLM stars and BDs with outflows, the main conclusion still reamins: the formation of VLM stars and BDs can be understood as a downsized version of low-mass stars.

\subsection{Revealing the true nature of Par-Lup3-4}
Par-Lup3-4 is a complex object that may help to understand the formation of VLM stars and BDs. Previous SED fitting \citep{Huelamo10-1} was not accurate because of the degeneracies between age and distance. The last Gaia data release broke the degeneracy by providing a precise distance value. Our SED fitting was underluminous in the optical, and this can best be explained with two main possibilities: either the source is younger than expected, or it is a binary. 
The presence of a bipolar molecular outflow, which is more common in Class 0 and I sources, together with the opening angle of the outflow points to a younger nature than Class II, and this is in agreement with one of the modeling results. 

We detect a stream of gas perpendicular to the detected bipolar outflow, whose origin is puzzling. We discussed four scenarios that might explain this. The possibility that this is a secondary molecular outflow, which would mean that ParLup 3-4 might be a binary, would agree with the results obtained in the SED fitting analysis.

Deep ALMA observations with higher resolution that point at the base of this possible secondary outflow may help to distinguish between outflowing material and cloud contamination, and might also help resolve the continuum emission of a possible binary source.
Additionally, ACA and Total Power ALMA observations are required to recover the larger spatial scales, in an attempt to detect the full extent of the outflow.

\section{Conclusions}
Par-Lup3-4 is a very low-mass protostar located in the Lupus 3 cloud. It has an edge-on disk with an optical jet. We observed Par-Lup3-4 using ALMA Bands 6 and 7, and we detected continuum and gas emission for three molecular lines (CO\,J=2-1, CO\,J=3-2 and $^{13}$CO\,J=3-2). These observations revealed for the first time the faint base of a molecular outflow and the cavity walls associated with this source, and a rotation pattern is seen with $^{13}$CO near location of the continuum source. The main results from this work are listed below. 

\begin{itemize}
\item The dust disk is faint and unresolved. The lower limit of the dust disk mass is 0.28 M$\mathrm{_{\oplus}}$.

\item The SED of Par-Lup3-4 can only be fit with models including a stellar radius 2 R$_{\odot}$, which is far from the 1.1 R$_{\odot}$ value derived in evolutionary models at the age of 2 Myr. We suggest that this radius value, and therefore a higher stellar luminosity, can be explained if Par-Lup3-4 is a close binary system. 
\item The average extent of the outflow is $\sim$300 au, which is a relatively short length for an outflow in the very low-mass regime. The outflow mass is found to be $\sim$10$^{-6}$M$_{\odot}$ , and the maximum outflow velocity we derive is 6 km/s.  We may not be observing the full extent of the outflow, and as a consequence, a portion of the outflow mass may be also missed. Our observations place lower limits on these outflow quantities. 
\item We detected a secondary structure that extends from the location of Par-Lup3-4 to the northeast and perpendicular to the primary molecular outflow that we report here. This might be a second outflow from Par-Lup3-4, suggesting that the source is a binary, or it may come from another nearby source. However, cloud contamination and a stream of infalling and rotating foreground material from the envelope cannot be discarded.

\end{itemize}

After measuring the particular properties of this VLM star, including the outflow length, mass, and maximum velocity, we compared our results with the predictions for VLM star and BD formation theories and found that they are all consistent with the formation of Par-Lup3-4 as a scaled-down version of low-mass star formation, as expected.

\begin{acknowledgements}
We thank the referee for their thorough review and comments that helped improve the quality of this manuscript. This work makes use of the following ALMA data:  ADS/JAO.ALMA$\#$2015.1.00512.S and ADS/JAO.ALMA$\#$2017.1.01401.S. ALMA is a partnership of ESO (representing its member states), NSF (USA) and NINS (Japan), together with NRC (Canada), MOST and ASIAA (Taiwan), and KASI (Republic of Korea), in cooperation with the Republic of Chile. The Joint ALMA Observatory is operated by ESO, AUI/NRAO and NAOJ. \\
A.S-M, and  M.R.S  acknowledge support from the "Iniciativa Cient\'ifica Milenio" via N\'ucleo Milenio de Formaci\'on Planetaria.
NH acknowledges finantial support from the Spanish State Research Agency (AEI) Project No. ESP2017-87676-C5-1-R and from project No. MDM-2017-0737 Unidad de Excelencia Mar\'ia de Ma\'eztu - Centro de Astrobiolog\'ia (CSIC-INTA).
NH and FC acknowledge support from the Faculty of the European Space Astronomy Centre (ESAC).
IdG is partially supported by MCIU-AEI (Spain) grant AYA2017-84390-C2-R (co-funded by FEDER).
\end{acknowledgements}

%
%

\bibliographystyle{aa} 
\bibliography{bibliografiav3} 
\begin{appendix} 
\section{Additional figures}
\begin{figure*}
\includegraphics{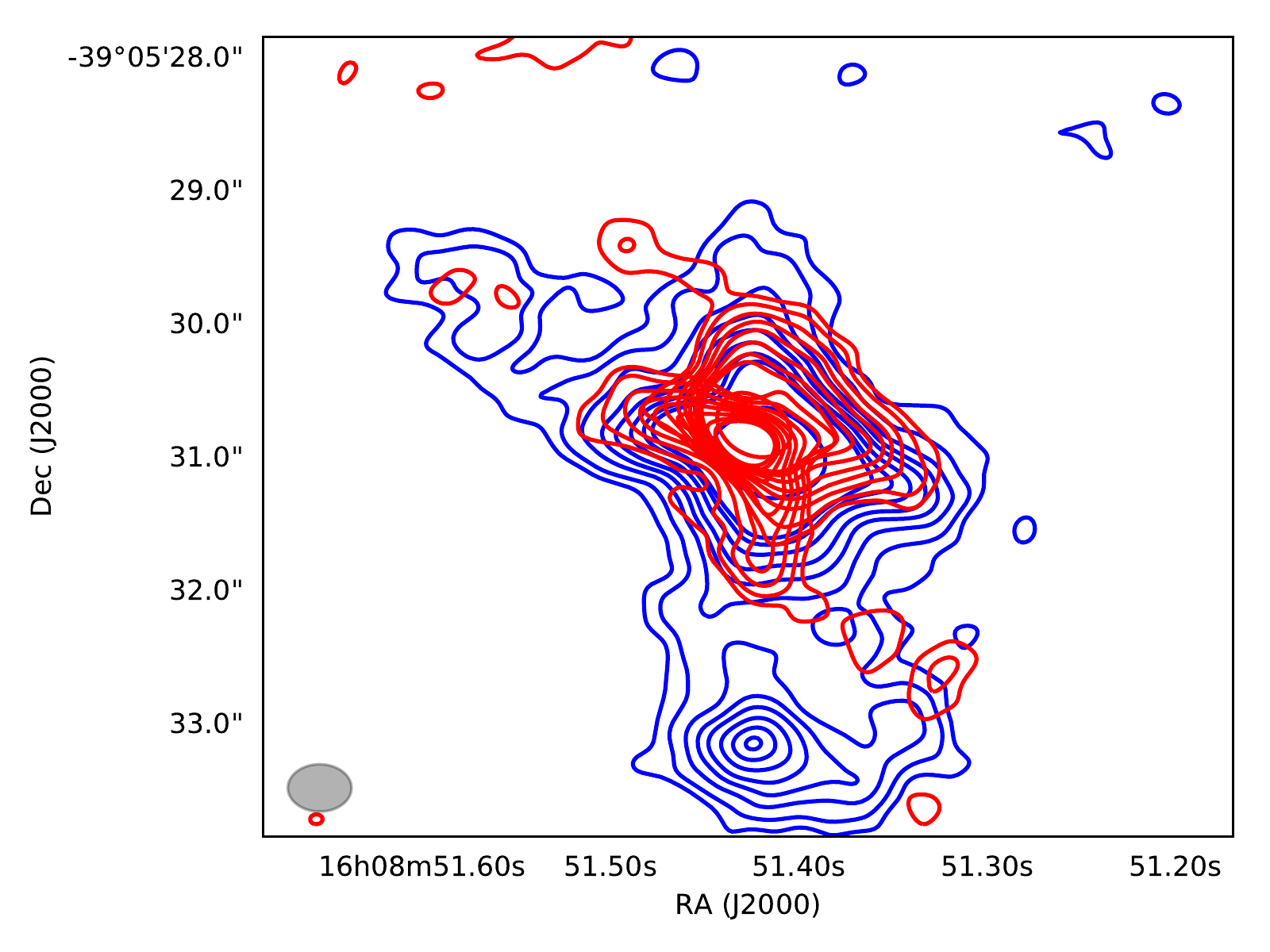}
      \caption{CO(3-2) integrated ALMA map. Blue contours show blueshifted emission between -1.6 to 2.36 km/s and red contours show redshifted emission between [4.12 km/s, 5.88 km/s] and [6.76 km/s, 9.84 km/s].  Contour levels are 3, 5, 8, 11, 14, 18, 22, 25, and 50 times the rms. The beam size is represented by a gray ellipse in the bottom left panel.} \label{red_blue_b7}
\end{figure*} 

\newpage
\newgeometry{inner=2.5cm, outer=2.5cm}
\begin{figure*}
\includegraphics[height=0.95\textheight]{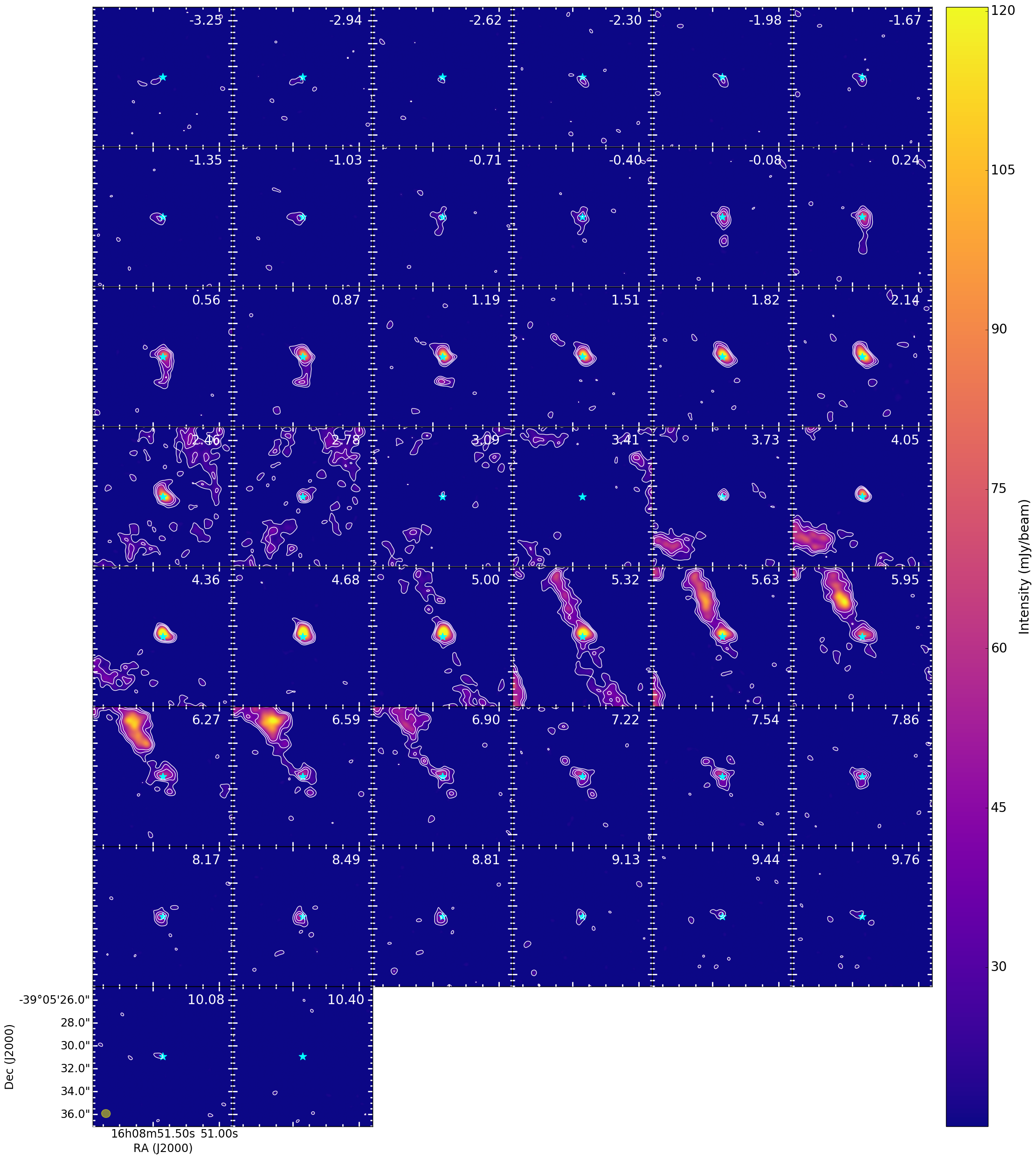}
      \caption[ALMA CO(2-1) channel emission map of Par-Lup3-4]{Zoom out CO(2-1) channel emission maps toward Par-Lup3-4. Contour levels are 3, 5, and 7 times the rms. The possible secondary outflow is between velocities 5.6 to 7.2 km/s from the center to the northeast of the image. The cyan star marks the position of the peak intensity of the continuum image. These maps have been obtained using a robust parameter equal to 1. The beam size is represented by a yellow ellipse in the bottom left panel. The LSR velocity in km/s is indicated in the top right corner of each image.} \label{channel_map_CO2-1}
\end{figure*} 
\restoregeometry

\begin{figure*}
\includegraphics{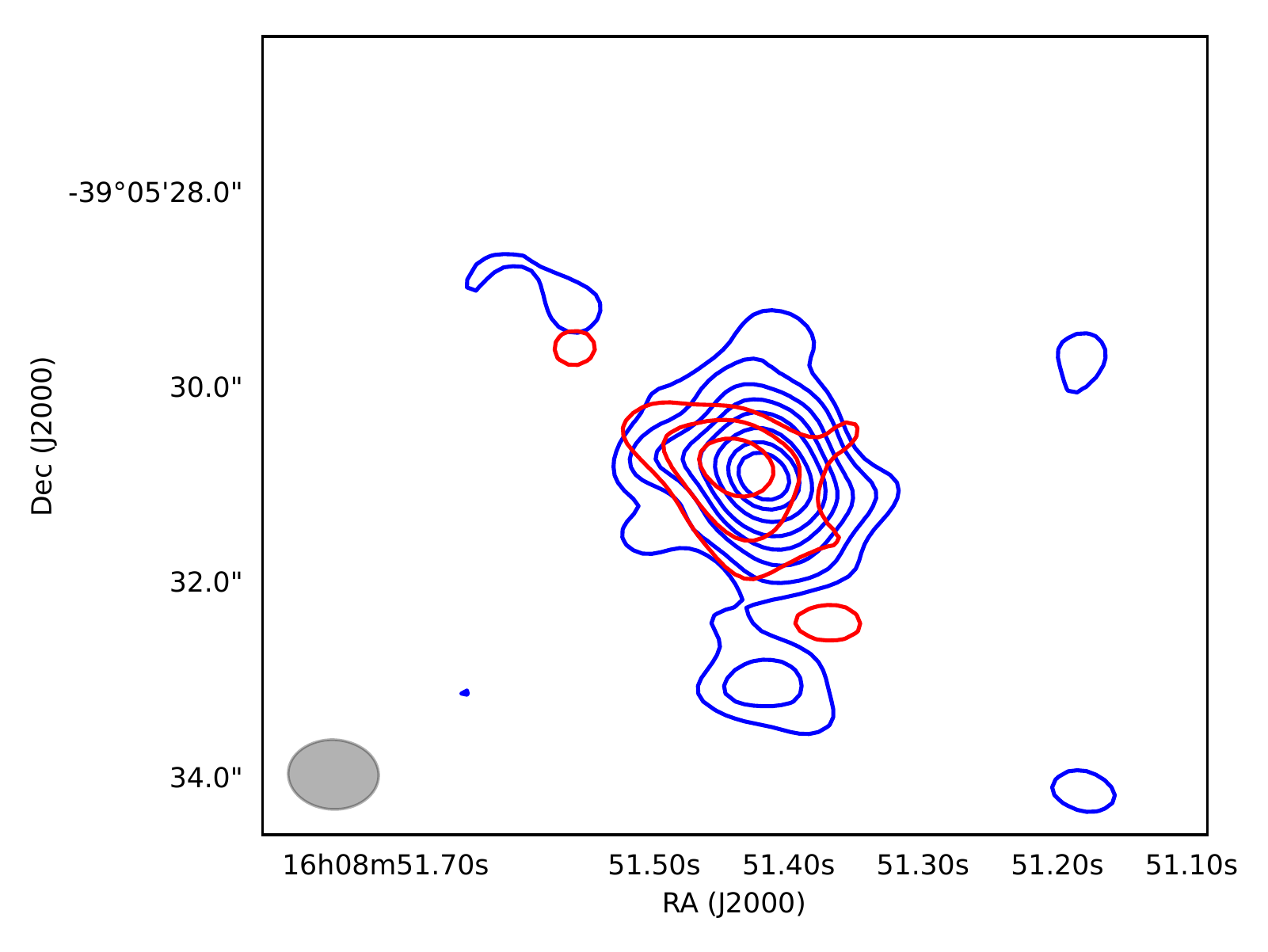}
      \caption{CO(2-1) integrated ALMA map. Blue contours show blueshifted emission between -2.9 to 2.14 km/s, and red contours show redshifted emission between 6.90 km/s and 10.0 km/s. Contour levels are 3, 5, 8, 11, 14, 18, 22, and 25  times the rms. The beam size is represented by a gray ellipse in the bottom left panel.} \label{red_blue_b6}
\end{figure*} 

\begin{figure*}[ht]
\includegraphics[width=\textwidth]{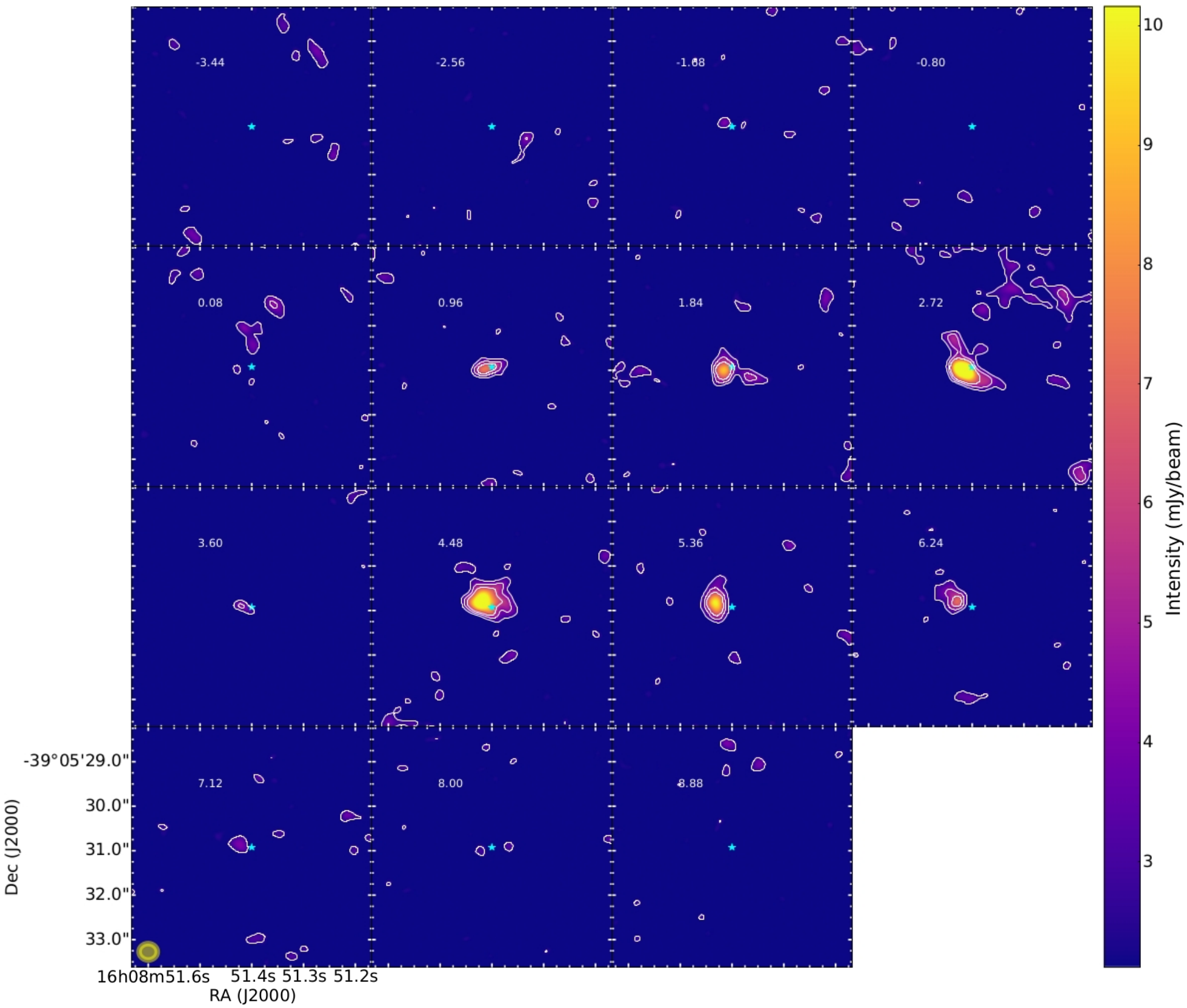}
\caption[ALMA $^{13}$CO(3-2) channel emission map of Par-Lup3-4]{\label{channel_map_13CO} Zoom-out $^{13}$CO channel maps toward Par-Lup3-4 using a robust value of 2. The velocity of the channels is shown in the LSR frame in km/s. Channels are binned to 0.88 km/s. All maps share the same linear color scale.  White contour levels are 3, 5, and 7 times the rms. The cyan star represents the position of the peak intensity in the continuum image.} 
\end{figure*}

\begin{figure*}
\includegraphics[width=0.5\textwidth]{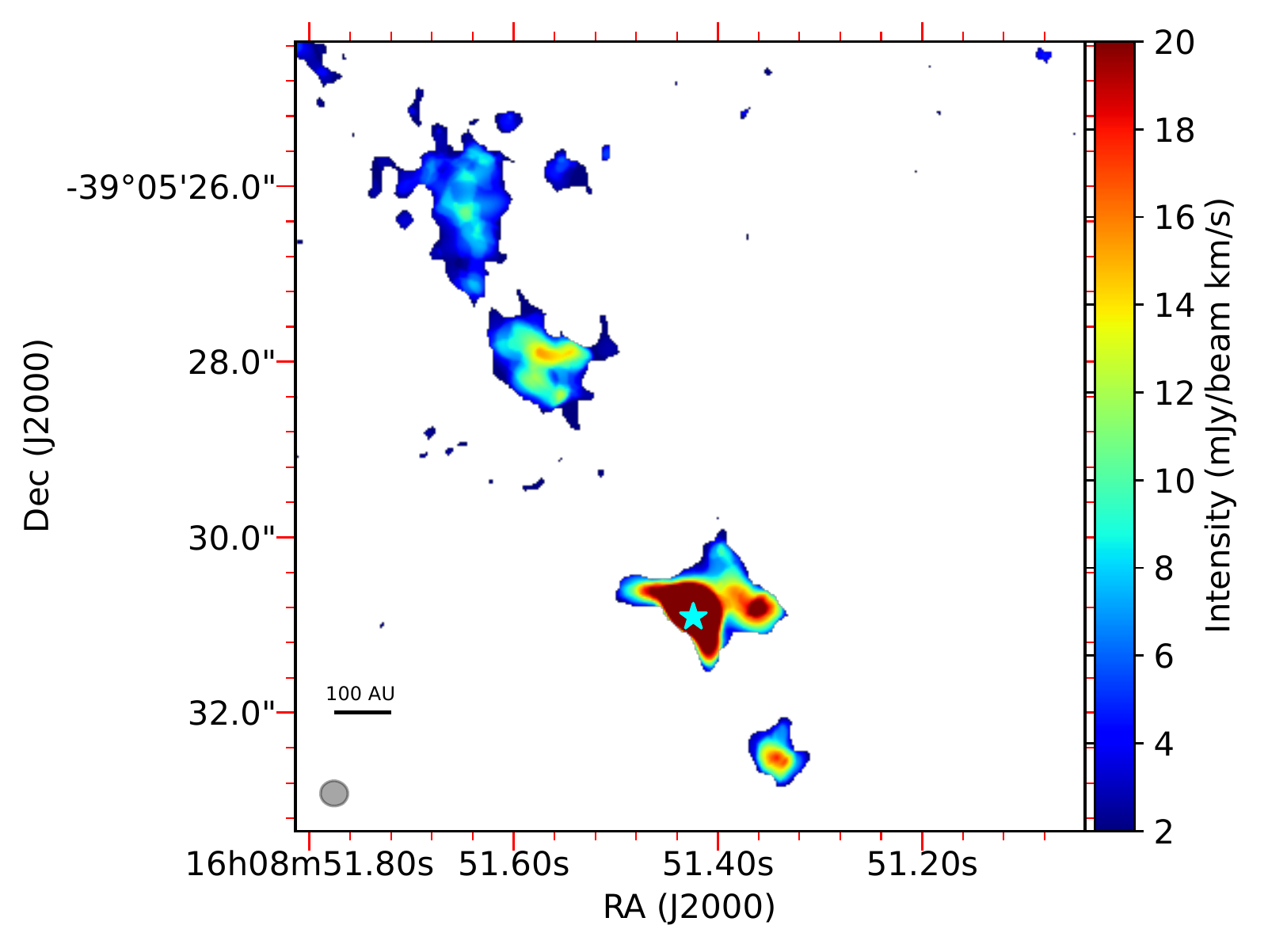}
\includegraphics[width=0.5\textwidth]{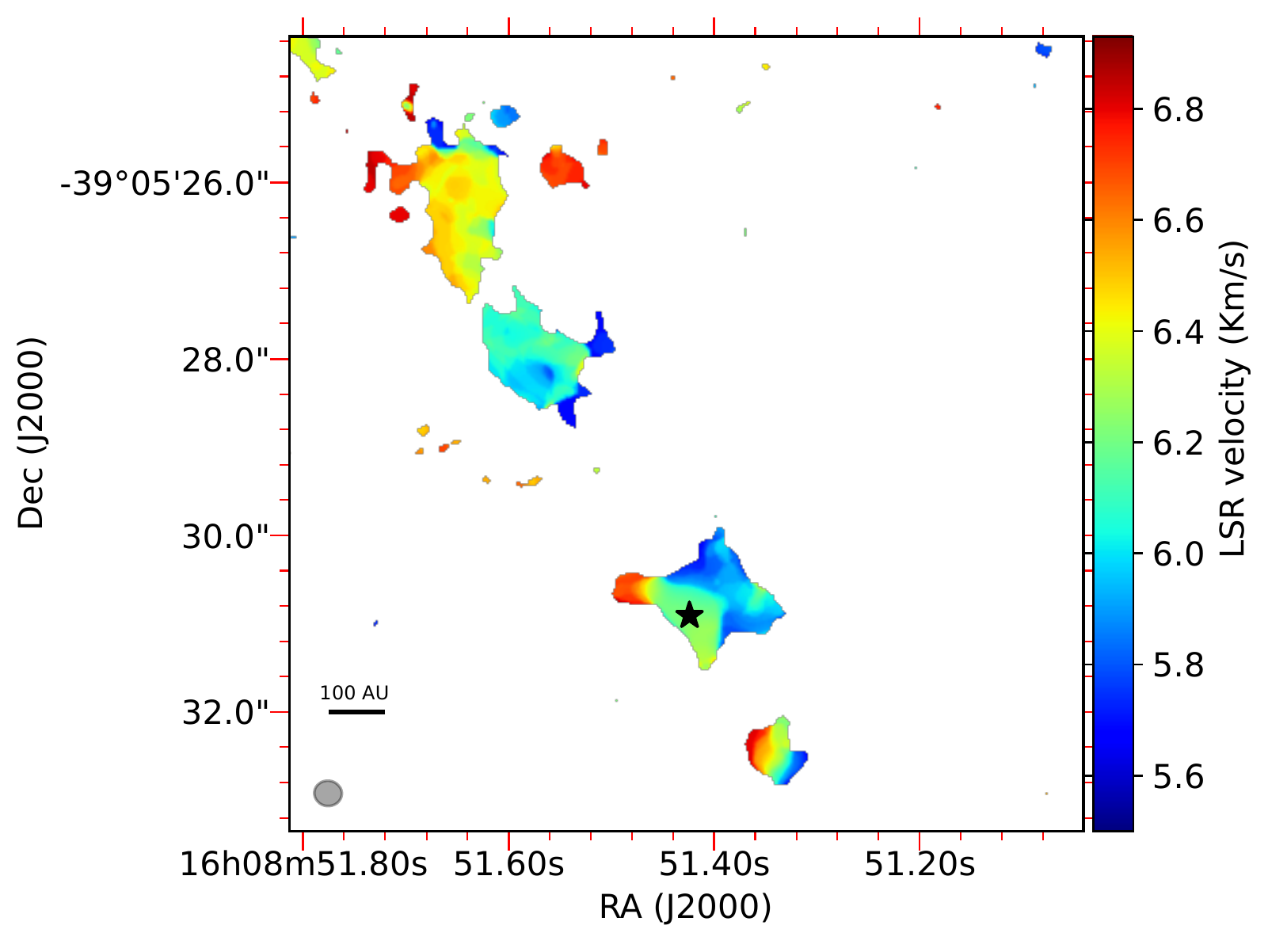}
\caption{Left panel: CO(3-2) integrated ALMA map. Right pannel: CO(3-2) ALMA velocity map. The beam size is represented by the gray ellipse in the bottom left corner. Only pixel values above 5$\sigma$ are included in the two panels. We used a uv range >50 k$\lambda$ to remove the extended emission. The cyan and black star marks the position of the peak intensity in the continuum image}\label{fig:CO_momentos_0_1_outflow_filtrado}
\end{figure*}

\begin{figure*}
\includegraphics[width=\textwidth]{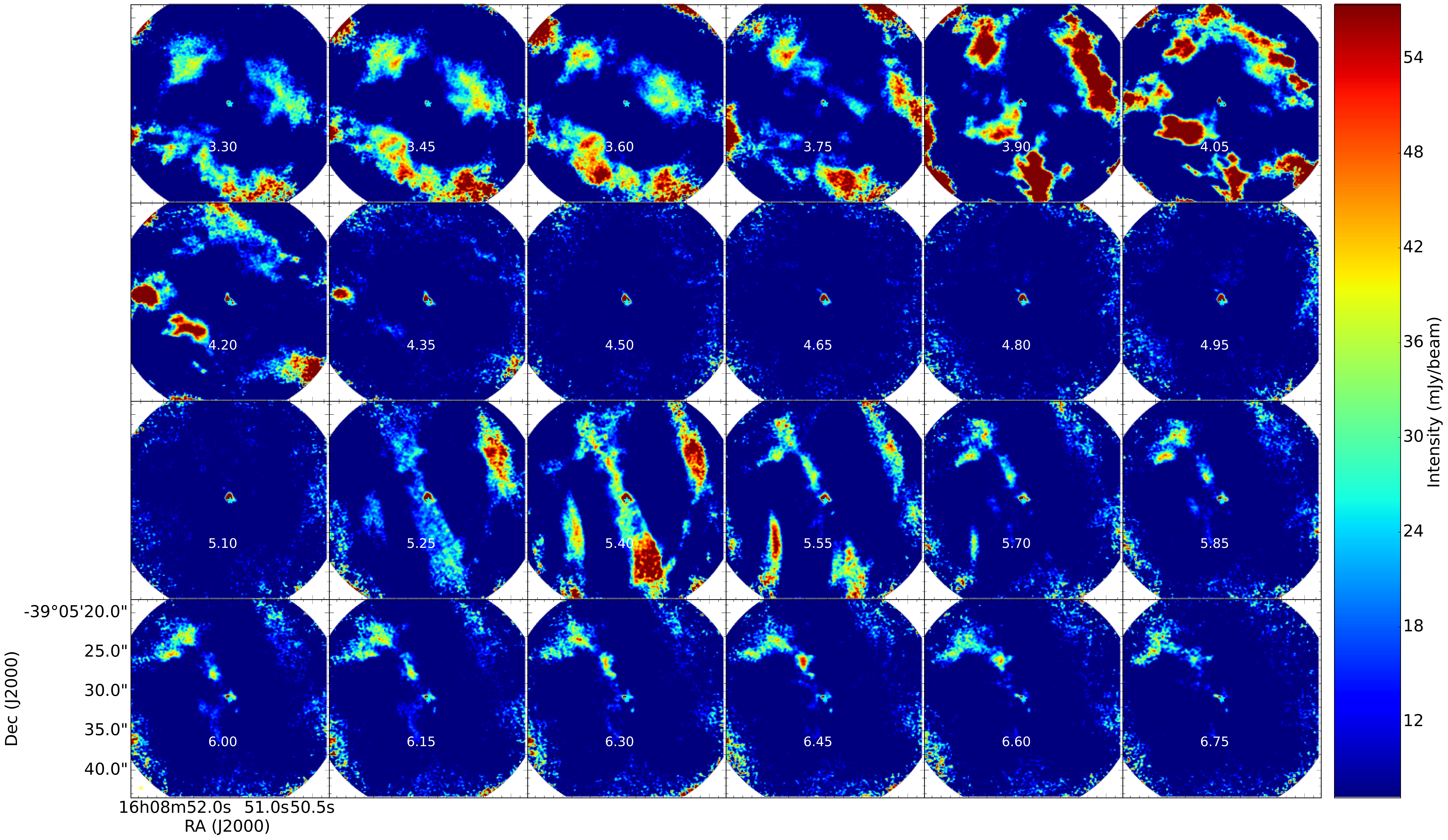}
      \caption[ALMA $^{13}$CO(3-2) channel emission map of the possible secondary outflow of Par-Lup3-4]{\label{channel_map_nube} CO(3-2) channel maps toward Par-Lup3-4 and the second outflow. The velocity of the channels is shown in the LSR frame in km/s. All maps share the same linear color scale.  the cyan star shows the position of the peak intensity in the continuum image.}
\end{figure*}

\end{appendix}

\end{document}